\documentclass[aps,twocolumn,pra,superscriptaddress,nofootinbib,amsmath,amssymb,floats,floatfix,english]{revtex4-1}
\usepackage{polski}
\usepackage[utf8]{inputenc}
\usepackage[T1]{fontenc}
\usepackage{amsmath}
\usepackage{amssymb}
\usepackage{ifthen}
\usepackage{graphicx}
\usepackage{times}
\usepackage{babel}
\usepackage{color}
\usepackage{bm}
\usepackage{pstricks}
\usepackage{ulem}

\DeclareMathSizes{7}{7}{5}{4} % \scriptsize is 7 pt
\DeclareMathSizes{6}{6}{3}{2} %"\smaller"
\DeclareMathSizes{5}{5}{3}{2} % \tiny is 5 pt

\DeclareMathSizes{5}{5}{2}{2}

\newcommand{\op}[1]{\widehat{#1}}
\newcommand{\dagop}[1]{\widehat{#1}^{\dagger}}

\newcommand{\mc}[1]{{\mathcal{#1}}}

\newcommand{\wb}[1]{{\overline{#1}}}

\newcommand{\nonu}{\nonumber}

\DeclareMathOperator{\TR}{Tr}

\newlength{\templength}

\newenvironment{eq}{\begin{equation}}{\end{equation}}

\newcommand{\eqn}[1]{(\ref{#1})}

\renewcommand{\eq}[2]{\begin{equation}\label{#1}#2\end{equation}}
\newcommand{\eqa}[2]{\begin{eqnarray}#2\label{#1}\end{eqnarray}}

\usepackage{lmodern}

\newcommand{\ve}{\varepsilon}
\begin{document}
\title{Mesoscopic density grains in the 1d interacting Bose gas from the exact Yang-Yang solution
}

\author{J. Pietraszewicz}
\affiliation{Institute of Physics, Polish Academy of Sciences, Aleja Lotnik\'ow
32/46, 02-668 Warsaw, Poland}
\email{pietras@ifpan.edu.pl, deuar@ifpan.edu.pl}

\author{P. Deuar}
\affiliation{Institute of Physics, Polish Academy of Sciences, Aleja Lotnik\'ow
32/46, 02-668 Warsaw, Poland}

\date{\today}

\begin{abstract}
Number fluctuations in a
one-dimensional Bose gas consist of contributions from many smaller independent localized fluctuations, the density grains.
We have derived a set of extended integral equations from the Yang-Yang solution for finite temperature that exactly determine all higher order moments of number fluctuations.
These moments are closely related to the statistics of the localized (but not zero-range) density grains.
We directly calculate the mean occupation of these fluctuations, and the variance, skewness, and kurtosis of their distribution across the whole parameter space of the gas. 
Findings include: 
Large mesoscopic density grains with a fat-tailed distribution in the thermal quasicondensate of the dilute gas and in the nonperturbative quantum turbulent regime;
Regions of negative skewness and below-Gaussian kurtosis in a part of the fermionized gas, and 
an unexplained crossover region along $T\sim T_d/\gamma$;
The existence of a peak in the density-density correlation function at finite interparticle spacing. 
We relate these density grain statistics to measurable behavior such as the statistics of coarse imaging bins, and finite-size scaling of number fluctuations. We propose how to experimentally test the relationship between thermodynamically independent density grains and density concentrations visible in single shot images. 
\end{abstract}

\maketitle 

%%%%%%%%%%%%%%%%%%%%%%%%%%%%%%%%%%%%%%%%%%%%%%%%%%%%%%%%%%%%%%%%%%%%%%%%%%%%%%%%%%%%%%%%%%%%%%%%%%%%%%%%%%%%%%%%%%%%%%%%%%%%%%%%%%%%%%%%%%%%%%%%%%%%%%%%%%%%%%%%%%%%%%%%%%%%%%%%%%%%%%%%%%%%%%%%%%%%%%%%%%%%%%%%%%
\section{Introduction}
\label{INTRO} 

One of the many ways in which one-dimensional quantum Bose gases are intriguing is that while they have highly quantum matter wave behavior such as e.g. superfluidity, they exhibit no off-diagonal long range order such as a condensate.
Hence, they must be composed of more or less microscopic scale, 
localized domains or \emph{grains} that survive into the thermodynamic limit. Very visually striking demonstrations of such domains have been seen in experiment especially for multicomponent gases, where the value of the local pseudospin acts as a marker, and spin domain walls are also not uncommon \cite{Sadler06,Vinit13,De14}.
The characteristics of the domains in a single-component gas are a little more elusive, since such clear-cut local markers are not available. 
The most familiar domains in this case are phase grains -- known to occur in the quasicondensate regime. Experimentally, their existence can be inferred from the behavior of a cloud after expansion \cite{Petrov00,Dettmer01,Hellweg03,Gring12,Deuar16}. 
There is also a warmer regime with profuse spontaneous solitons \cite{Karpiuk12,Karpiuk15,Gawryluk17,Nowicki17} separating phase and density domains. 
 Overall, due to the lack of a condensate, phase grains of various sorts are to be expected in other regimes as well. 

Density granularity, in the sense of fluctuating local-scale structures, is 
also expected due to the localized behavior of the density-density correlation function $g^{(2)}(z)$  
in all regimes of the 1d Bose gas \cite{Sykes08,Deuar09,Wang13,Nandani16}.
The statistics and qualities of these \emph{density grains} is what we reveal here, having developed a technique to study them in the exact quantum solution for the thermodynamic limit. 

The uniform one-component Bose gas in the thermodynamic limit has an exact solution at $T=0$ due to Lieb and Liniger \cite{Lieb63,Lieb63b}, extended to the $T>0$ case by Yang and Yang \cite{Yang69}. The basic quantities calculated from this solution, such as density, pressure, energy per particle are intensive system-averaged quantities and they do not meaningfully relate to localized elements of the gas. Extracting other observables from the exact solution is usually nontrivial. In a broad effort, a selection of microscopic properties have been found already. The local among them are the  density-density correlation $g^{(2)}(0)$ \cite{Kheruntsyan03,Kheruntsyan05,Wang13}, 
$g^{(3)}(0)$ \cite{Cheianov06,Kormos11}, and some other quantities \cite{Kormos10,Piroli15,Nandani16}.
The second kind are two-body:  dynamic and static structure factors \cite{Panfil14,Panfil13,deNardis16}, response functions \cite{Cherny06},
and two-body correlations \cite{Sykes08,Deuar09,Golovach09,Kozlowski11,Kozlowski11b,Patu13,Klumper14,Nandani16} in some physical limits.
In particular, the $k=0$ static structure factor (denoted $S_0$) gives information about occupation fluctuations in imaging bins, in the limit of coarse-enough bins, and has been used in this capacity to interpret experimental measurements \cite{Esteve06,Sanner10,Muller10,Armijo10,Jacqmin11,Armijo11,Armijo12,Bisset13}.
The link to physics on the localized but not zero-range scale is evident. Furthermore, it can be obtained from an appropriate thermodynamic derivative of the total number of particles in the system, $N$ \cite{Armijo10}. 

What we found at first was an alternative method to calculate $S_0$  from the exact Yang-Yang solution. This method quite naturally lends itself to a whole hierarchy extensions that allow one to calculate higher moments of $N$.  
Furthermore, by considering a natural criterion of independence for density fluctuations --- that the variance of the total atom number be the sum of variances of the \emph{independent} fluctuations --- we find a way to relate the system-size scaling of some extensive quantities to the behavior of mesoscopic physical elements in the gas - the density grains. 

In consequence, we are able to determine the statistics of the independent density fluctuations in the gas across the whole range  of physical regimes: the quasicondensate, the dilute degenerate gas including the quantum turbulent regime, the classical gas, and the strongly interacting fermionized regime, as well as all the crossovers. 
We explicitly obtain the mean number of correlated particles in a fluctuation, how Poissonian/sub-Poissonian the occupations are, the skewness of their distribution, and its kurtosis. 
Interestingly, neither skewness nor kurtosis could be obtained merely from considering known results on two-body correlation functions. 
We note both tail-heavy distributions in the quasicondensate, as well as platykurtic and/or negatively skewed distributions in the nonzero temperature fermionized gas.
To our best knowledge, exact quantum results for mesoscopic objects that can involve many particles have not been studied earlier in this system.

The paper is organized as follows:
Sec.~\ref{SOL} introduces the system, its basic parameters, and the Yang-Yang solution, as well as a visualization of the iteration procedure used. 
The Poissonian/sub-Poissonian statistics of the density grains and the underlying method we use to obtain it is derived in Sec.~\ref{DGR}, while Sec.~\ref{GRAINS} explains how the mean number of particles partaking in such independent fluctuations can be ascertained. Having done this, the skewness and kurtosis of the density grain distribution are  found in Secs.~\ref{SKEW} and~\ref{KURT}, respectively. 
We make physical comments about the main results obtained (presented in Figs.~\ref{fig:uG}-\ref{fig:kurt}) as we go along. Suggestions on how these quantities can be experimentally observed are given in Sec.~\ref{OBS}. We conclude in Section~\ref{CONCLUSIONS}.

%%%%%%%%%%%%%%%%%%%%%%%%%%%%%%%%%%%%%%%%%%%%%%%%%%%%%%%%%%%%%%%%%%%%%%%%%%%%%%%%%%%%%%%%%%%%%%%%%%%%%%%%%%%%%%%%%%%%%%%%%%%%%%%%%%%%%%%%%%%%%%%%%%%%%%%%%%%%%%%%%%%%%%%%%%%%%%%%%%%%%%%%%%%%%%%%%%%%%%%%%%%%%%%%%%
\section{System and exact solution}
\label{SOL}

%%%%%%%%%%%%%%%%%%%%%%%%%%%%%%%%%%%%%%%%%%%%%%%%%%%%%%%%%%%%%%
\subsection{System and units} 
\label{UNITS}
We will be considering a uniform gas described by the basic ultracold atom Hamiltonian
\eq{H}{
\op{H} = \int dx \left\{\dagop{\Psi}(x)\left[-\frac{\hbar^2}{2m}\frac{d^2}{dx^2}\right]\op{\Psi}(x) + \frac{g}{2}\dagop{\Psi}(x)^2\,\op{\Psi}(x)^2 \right\}.
}
in a 1d box of length $L$, with 
\eq{N}{
\op{N} = \int dx\ \dagop{\Psi}(x)\op{\Psi}(x)
}
particles. 
Imagine a section of the gas which stays in diffusive contact with more cloud (at the ends, or in higher transverse excited states).
Then, the expected equilibrium ensemble will be a grand canonical one with chemical potential $\mu$. 
This is also the ensemble assumed in the Yang-Yang exact solution for the nonzero temperature gas in the thermodynamic limit \cite{Yang69}.

First, a minor but often confusing matter of variables. The Lieb-Liniger \cite{Lieb63,Lieb63b} and Yang-Yang Bethe ansatz solutions are given in units of $\hbar=2m=1=k_B$, with a coupling strength $c$ and chemical potential $A$.
Here, we will use the $\hbar=m=1$ units more familiar in ultracold atoms to avoid a ``factor of two'' curse when comparing to other cold atom work.
The distance units (for $k$, $L$, etc) will stay the same as in the Lieb-Liniger description. 
These steps result in the following relationship:  the mass and time units are twice as large in Lieb-Liniger than standard cold atom fare
while energy, $T$, and coupling $g$ have units that are two times smaller in Lieb-Liniger. 
Therefore, if we denote quantities appearing in the Lieb-Liniger, Yang-Yang, and related papers with subscript ${}_{LL}$, and the standard 
 $\hbar=m=k_B=1$ case without subscripts, one has:
$E=\frac{1}{2}E_{LL}$, $\mu=A_{LL}/2$, $T=T_{LL}/2$, $k=k_{LL}$ and $g=c_{LL}$.

%%%%%%%%%%%%%%%%%%%%%%%%%%%%%%%%%%%%%%%%%%%%%%%%%%%%%%%%%%%%%%
\subsection{The plain equations} 
\label{EQS}
To use the Yang-Yang Bethe ansatz solution \cite{Yang69} in practice, one first solves the following integral equation:
\eq{ek-ieq}{
\ve(k) = -\mu + \frac{k^2}{2} -\frac{gT}{\pi} \int_{-\infty}^{\infty} \frac{dq\,\log\left[1+e^{-\ve(q)/T}\right]}{g^2+(k-q)^2}.
}
The spectrum-like quantity $\ve(k)$ depends on wavevectors~ $k$.\footnote{Note that we also rescale this quantity to $\hbar=m=1$ units so that $\ve=\epsilon_{LL}/2$ compared to the quantity $\epsilon_{LL}$ used in \cite{Yang69}.}
A continuum of wavevectors is considered, since the theory is for the thermodynamic limit of sufficiently large~ $L$.
The equation \eqn{ek-ieq}  is usually solved by iteration, starting with the free particle form
\eq{iter0}{
\ve^{(0)}(k) = -\mu + \frac{k^2}{2}.
}
Subsequently, the following Fredholm integral equation
\eq{rhok-ieq}{
2\pi\rho(k)\left[1+e^{\ve(k)/T}\right] = 
1 + 2g \int_{-\infty}^{\infty} \frac{dq\ \rho(q)}{g^2+(k-q)^2}.
}
is solved for a function $\rho(k)$ which gives the density of occupied quasimomenta.
This step is also done by iteration, starting from
\eq{iter0-rho}{
\rho^{(0)}(k) = \frac{1}{2\pi\left[1+e^{\ve(k)/T}\right]}.
}

\begin{figure}
\begin{center}
\includegraphics[width=\columnwidth]{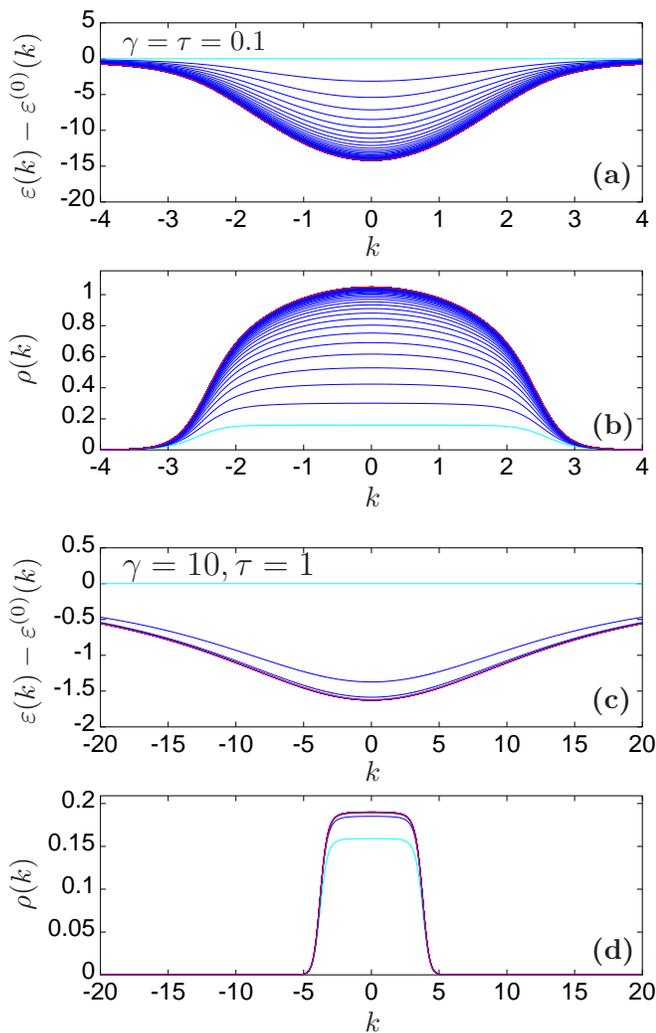}
\end{center}
\caption{
Integration kernels and iteration details for two characteristic cases. In the top two panels: a weak quasicondensate with $\gamma=\tau=0.1$. In the bottom two panels: a moderately high temperature fermionized gas with $\gamma=10$, $\tau=1$. 
The top plot in each pair shows the shift of the spectrum-like quantity $\ve(k)$ from the free particle form, $\delta\ve=\ve(k)-(\frac{k^2}{2}-\mu) = \ve(k)-\ve^{(0)}(k)$, while the lower shows the quasimomentum occupation density $\rho(k)$.
The cyan lines are the initial estimates $\delta\ve^{(0)}(k)=0$ and $\rho^{(0)}(k)$. The blue lines are the results of subsequent iterations, and the red lines are the final self-consistent solutions of the integral equations.
}
\label{fig:vars}
\end{figure}

Figure \ref{fig:vars} shows some typical behavior during this iterative procedure.
Convergence can be fast or time consuming depending on the physical regime. The quasicondensate regime $\gamma\ll1$, $\tau\ll\sqrt{\gamma}$ has particularly slow convergence,
while convergence is very fast in the strongly fermionized regime when $\gamma\gg1$.

Using the above solution, the following physical quantities are obtained in a straightforward manner \cite{Yang69}:\\
the density
\eq{nsol}{
n=\frac{N}{L} = \frac{\langle\op{N}\rangle}{L} = \int_{-\infty}^{\infty} \rho(k) dk,
}
the energy per particle $E/N$:\\
\eq{Esol}{
\mc{E}=\frac{E}{N}=\frac{\langle\op{H}\rangle}{\langle\op{N}\rangle} = \frac{1}{2n}\int_{-\infty}^{\infty} \rho(k) k^2 dk,
}
the entropy $S$, and the pressure $P$: \\
\eq{Psol}{
P= \frac{T}{2\pi}\int_{-\infty}^{\infty} dk\, \log\left[1+e^{-\ve(k)/T}\right].
}
In order to obtain another important observable characterizing the system, namely the local density-density correlation function \cite{Kheruntsyan03,Kheruntsyan05}, one can further use the  Hellman-Feynmann  theorem as follows:
\eq{g20sol}{
g^{(2)}(0) = \frac{1}{n^2}\langle\dagop{\Psi}(x)^2\op{\Psi}(x)^2\rangle = -\frac{1}{n^2}\left(\frac{\partial P}{\partial g}\right)_{\mu,T}.
}
Intense numerical calculations based directly on the Bethe ansatz have also been used to obtain dynamic structure factors and correlations \cite{Panfil13,Panfil14,deNardis16}.

%%%%%%%%%%%%%%%%%%%%%%%%%%%%%%%%%%%%%%%%%%%%%%%%%%%%%%%%%%%%%%
\subsection{Parameter space and scaling} 
\label{GAMTAU}
Now in the thermodynamic limit of a uniform gas with density $n=N/L$ there are only two essential parameters:\\
an interaction strength
\eq{gam}{
\gamma = \frac{mg}{\hbar^2 n}
}
and a relative temperature 
\eq{tau}{
\tau = \frac{2mk_BT}{\hbar^2n^2} = \frac{4\pi T}{T_d}.
}
Here, $T_d=\hbar^2n^2/2m$ is the usual ideal gas degeneracy temperature, while the $4\pi$ factor has often been used in the context of density fluctuations.

The Yang-Yang Bethe ansatz description is given in terms of three input parameters $g, T, \mu$, and the density is 
then a function of these as per $n(\mu,T,g)$.
Due to this, a pair of physical parameters $(\gamma,\tau)$ corresponds to continuous families of gases in the thermodynamic limit
and we will use them to describe the parameter space. 

Typically, the main regimes of density fluctuations have been classified using the local microscopic quantity $g^{(2)}(0)$ \cite{Kheruntsyan03}:
\begin{itemize}
\item A fermionized regime ($\gamma\gg1$ and $\tau\ll\gamma^2$) with strong antibunching $g^{(2)}(0)\ll1$. 
An intriguing high-temperature fermionized regime occurs for $1\ll\tau\ll\gamma^2$.
\item A quasicondensate regime ($\gamma\ll1$ and $\tau\ll\sqrt{\gamma}$) with small density fluctuations $|g^{(2)}(0)-1|\ll1$. This regime is further distinguished into a thermal fluctuation dominated region with $g^{(2)}(0)>1$ for $\tau\gtrsim\gamma$ and a quantum fluctuation dominated region  with $g^{(2)}(0)<1$ for $\tau\lesssim\gamma$.
\item A decoherent quantum regime ($\sqrt{\gamma}\ll\tau\ll1$) in which density fluctuations are large ($g^{(2)}(0)-1\sim\mc{O}(1)$) but the system is still quantum degenerate.
\item A classical particle-like regime for higher temperatures in which $g^{(2)}(0)\approx2$.
\end{itemize}
Two features in the dilute gas at intermediate temperature are also worth noting for later comparison: 
\begin{itemize}
\item A regime with frequently occurring thermally activated solitons in the range $\tau\sim (0.5-2.5)\sqrt{\gamma}$ \cite{Karpiuk12,Nowicki17}.
\item The crossover around $\tau\sim(3-4)\sqrt{\gamma}$ which occurs when $\mu$ changes sign. For positive $\mu$, the system is Bogoliubov-like, while for $\mu<0$ the description is rather of the Hartree-Fock type, as studied in \cite{Henkel17}.
\end{itemize}

In this paper, we will choose $T$ to be the scaling parameter, and use
the value $T=1$ for numerical calculations. 
Since $n=(1/\hbar)\sqrt{2mk_BT/\tau}$ and $g=n\gamma\hbar^2/m$, the whole family of solutions has the following scalings:
\eqa{sca}{
\frac{n}{\sqrt{T}} = n|_{T=1}; \qquad 
\frac{g}{\sqrt{T}} = g|_{T=1}; \qquad 
\frac{\mc{E}}{T} = \mc{E}|_{T=1}\nonu\\
\frac{\mu}{T} = \mu|_{T=1}; \qquad
\frac{P}{T^{3/2}} = P|_{T=1};\qquad
\dots \quad
} 
The energy and chemical potential scalings follow from the fact that they have the same units as 
temperature, while the thermodynamical relation $dP~=~(S/L)dT~+~n\,d\mu$ indicates the scaling of pressure.

In calculations aiming for a given parameter pair $(\gamma,\tau)$,  sought-after values of $n$ and $g$ are uniquely determined after choosing $T=1$. However, we  need to find the $\mu$ that gives the appropriate value of the density. This is a numerical inverse problem that can be solved by standard numerical techniques once we are able to evaluate $n(\mu,g,T)$. Typically, it  takes from a few to a few tens of steps to get 4-5 significant digits of accuracy in $n$, largely irrespective of the physical regime.

%%%%%%%%%%%%%%%%%%%%%%%%%%%%%%%%%%%%%%%%%%%%%%%%%%%%%%%%%%%%%%%%%%%%%%%%%%%%%%%%%%%%%%%%%%%%%%%%%%%%%%%%%%%%%%%%%%%%%%%%%%%%%%%%%%%%%%%%%%%%%%%%%%%%%%%%%%%%%%%%%%%%%%%%%%%%%%%%%%%%%%%%%%%%%%%%%%%%%%%%%%%%%%%%%%
\section{Density grains and their statistics}
\label{DGR}
%%%%%%%%%%%%%%%%%%%%%%%%%%%%%%%%%%%%%%%%%%%%%%%%%%%%%
\subsection{Coarse grained density fluctuations }
\label{UG}

The local pinpoint density-density correlation function $g^{(2)}(0)$ describes very small scale fluctuations. 
While it is a very important quantity for the theoretical description of the gas, it is not what is usually observed. That is because finite imaging resolution in a typical setup makes it inaccessible. 

Under typical conditions, the finite resolution of the imaging apparatus, $\Delta$, is comparable to or wider than the width of the $g^{(2)}(z)$ density correlation function
\eq{g2zdef}{
g^{(2)}(z) = \frac{1}{n^2}\langle\dagop{\Psi}(x)\dagop{\Psi}(x+z)\op{\Psi}(x+z)\op{\Psi}(x)\rangle.
}
This means that the statistics of the observed local fluctuations are in fact only the statistics of segments of the gas of length $\Delta$. 
This is the case in most contemporary experiments\footnote{Special setups with single-atom detection in He${}^*$ have been able to access $g^{(2)}(z)$ in detail, though \cite{Perrin07,Jaskula10,Kheruntsyan12,Dall13,Manning13}.}.
For observations with finite resolution, the results can be described by a sequence of bins of width $\Delta$, with observed occupations $N_j^{(\Delta)}$ in the $j$th bin. 
Ensemble averages such as $\langle N_j^{(\Delta)}\rangle = \langle N^{(\Delta)}\rangle$ will be independent of $j$ for a uniform system.
An in-depth experimental study of such \emph{coarse-grained} density fluctuations and their statistics has been carried out by the Palaiseau group \cite{Armijo10,Armijo11,Jacqmin11,Armijo12}. 

A fundamental statistical quantity in this regard is 
\eq{uGDelta}{
\frac{{\rm var} N^{(\Delta)}}{\langle N^{(\Delta)}\rangle},
}
which compares the bin occupations to Poissonian statistics. A value of unity indicates Poissonian variance, values above one super-Poissonian variance, and below one: sub-Poissonian. The last  can only occur at sufficiently low temperatures when quantum fluctuations dominate.
The quantity in \eqn{uGDelta} is evidently device-dependent when $\Delta$ is small, because
 some density correlations may occur between neigboring bins. 

Now instead of these imaging-limited bins, consider bin sizes $\mc{L}$ that are sufficiently large to have statistically independent occupations, but still $\mc{L}\ll L$. We then arrive at the following intensive thermodynamic quantity which describes coarse grained density fluctuations:
\eq{uG}{
S_0  = \frac{{\rm var} N}{N} = \frac{{\rm var} N^{(\mc{L})}}{\langle N^{(\mc{L})}\rangle}.
}
This is also known as the $k = 0$ static structure factor. 
The last equality in \eqn{uG} follows from the assumption of independence between the occupations $N^{(\mc{L})}_j$, because 
the variance of the total particle number $N$ grows as the sum of the variances of the individual independent contributions. 
$S_0$ depends neither on the ultimate size of the box $L$, nor on the bin size $\mc{L}$ provided the latter are sufficiently large. 
The matter of ``sufficiently large'' can be quantified by requiring that the density correlation function $g^{(2)}(z)$ decays to its background value of one when $z\ge\mc{L}$. 
We can see this relationship clearly by evaluating $S_0$ from \eqn{uG}, substituting $\op{N}=\int dx\,\dagop{\Psi}(x)\op{\Psi}(x)$ for the number of particles in the system, and comparing to \eqn{g2zdef}. The result is 
\eq{uGg2}{
S_0 = 1 + n\int_{\rm system} dz \left[g^{(2)}(z)-1\right].
}
If the integral is over a greater extent than the width of the bulge in the correlation function at low $z$, then it achieves its asymptotic value. 

From \eqn{uGg2} one readily sees several features:
$S_0\le 1$ only if there is antibiunching ($g^{(2)}(z)<1$), an effect that only occurs at temperatures low enough that quantum depletion becomes important. In the nondegenerate classical gas, we achieve the Poissonian shot noise limit $S_0\to1$, since then both $n$ and the range of $g^{(2)}(z)$, which is of the order of the thermal de Broglie wavelength $\Lambda_T$, become small. In the condensate, when $g^{(2)}(z)=1$, we again have a shot noise value of $S_0=1$, but for a different reason.

%%%%%%%%%%%%%%%%%%%%%%%%%%%%%%%%%%%%%%%%%%%%%%%%%%%%%
\subsection{Independent density grains}
\label{DGRAINS}
The quantity $S_0$ is also a descriptor of the typical independently occurring ``lumps'' of density, regardless of any considerations of an externally set bin size.

To see this, consider the following: if we have $p$ independently fluctuating fragments of the gas labeled by $j$, which we will dub \emph{density grains},  then
\eq{N=}{
N=\langle\op{N}\rangle=\sum_{j=1}^{p} \langle\op{\mc{N}}_j\rangle = p\wb{\mc{N}}.
}
We have denoted the number of particles in individual grains by $\op{\mc{N}}_j$ in italics, and a bar will be used to indicate averaging over grains. For example,
\eq{mcNbar1}{
\wb{\mc{N}} = \frac{1}{p}\sum_{j=1}^p\langle\op{\mc{N}}_j\rangle.
}
is the average grain occupation.  
In the thermodynamic limit with many grains $p\to\infty$, an average over independent grains in a single experimental realization will converge to the same value as the ensemble average. So that
\eq{mcNbar}{
\lim_{p\to\infty}\wb{f(\mc{N})} = \frac{1}{p}\sum_{j=1}^pf\left(\mc{N}_j\right)
}
also holds for general quantities $f$ that involve only measured grain occupations $\mc{N}_j$.

Consider now the second moment of the total atom number:
\eqa{N2mc}{
\langle \op{N}^2\rangle  &=& \sum_{jj'}\langle\op{\mc{N}}_j\op{\mc{N}}_{j'}\rangle = \sum_j\langle\op{\mc{N}}_j^2\rangle + \sum_{j\neq j'}\langle\op{\mc{N}}_j\rangle\langle\op{\mc{N}}_{j'}\rangle \nonu\\
&=& p \left[\wb{\mc{N}^2} + (p-1)\wb{\mc{N}}^2\right].
}
Hence, 
\eq{N2mc2}{
{\rm var} N = p\left[\wb{\mc{N}^2} -\wb{\mc{N}}^2\right] = p\,{\rm var}\mc{N}.
}
After all, we expect that the variance of a collection of independent random variables will be  the sum of their variances. That is what is often meant by \emph{independent}.
We can see also that
\eq{N2mc3}{
S_0 = \frac{{\rm var}N}{N} = \frac{{\rm var}\mc{N}}{\wb{\mc{N}}}
}
gives information about the statistics (sub-Poissonian, Poissonian, super-Poissonian) of the individual independent density grains in the gas. 
Interestingly, this information is given without reference to the separate  question of how large such density grains are (i.e. the actual value of $\wb{\mc{N}}$).  

The defining quality of what we will mean by a \emph{density grain} is that the local variances of independent grains are additive to the global variance.
One example are segments of sufficient length $\mc{L}$. These will become independent as boundary effects between them become negligible. That is, when $\mc{L}\ge\Delta_c$, and $\Delta_c$ is the bin size at which \eqn{uGDelta} begins to deviate from $S_0$. A different example would be fuzzy segments defined by an integral over a point spread function with width $\Delta\gtrsim\Delta_c$.
We will return to the matter of the density grain size in more detail in Sec.~\ref{GRAINS}.

%%%%%%%%%%%%%%%%%%%%%%%%%%%%%%%%%%%%%%%%%%%%%%%%%%%%%
\subsection{Calculating the density grain statistics}
\label{UGCALC}

%%%%%%%%%%%%%%%%%%%%%%%%%%%%%%%%%%%%%%%%%%%%%%
\subsubsection{Thermodynamic relation}
Following e.g. \cite{Armijo10}, consider the grand canonical ensemble with fixed $g$ and $T$.
Denoting the grand canonical partition function by $\mc{Z} = \TR[\op{Z}]$, where $\op{Z}=\exp\left[-\left(\op{H} - \mu \op{N}\right)/T\right]$, 
one has that 
\eq{wbNZ}{
\langle \op{N} \rangle = \frac{{\rm Tr}[\op{N}\op{Z}]}{\mc{Z}} = \frac{k_BT}{\mc{Z}}\,\left(\frac{\partial\mc{Z}}{\partial\mu}\right)_{g,T}.
}
Also,
\eq{wbN2Z}{
\langle \op{N}^2 \rangle = \frac{(k_BT)^2}{\mc{Z}}\,\left(\frac{\partial^2\mc{Z}}{\partial\mu^2}\right)_{g,T}.
}
Denoting the fluctuations as 
\eq{delN}{
\delta\op{N} = \op{N} - \langle\op{N}\rangle
}
one immediately has the well known thermodynamic relation
\eq{wbN2mu}{
\langle\delta \op{N}^2\rangle = {\rm var}N = k_BT\,\left(\frac{\partial\langle\op{N}\rangle}{\partial\mu}\right)_{g,T}.
}
Now using \eqn{uG} and $n=N/L=\langle N\rangle/L$ one obtains
\eq{uGu}{
S_0 = \frac{k_BT}{n}\,\left(\frac{\partial n}{\partial\mu}\right)_{g,T}.
}

%%%%%%%%%%%%%%%%%%%%%%%%%%%%%%%%%%%%%%%%%%%%%%%
\subsubsection{Basic approach}
The most straightforward way to calculate $S_0$  is to proceed by converting the partial derivative in \eqn{uGu} to finite differences. This is  what has usually been done for the microscopic local density fluctuations $g^{(2)}(0)$ using the prescription \eqn{g20sol} \ \cite{Kheruntsyan05}\footnote{The formula using $P$ from \cite{Kheruntsyan05} is more convenient in practice than the better known one using free energy that was derived earlier in \cite{Kheruntsyan03}.}. 
In that approach,  a small shift $\Delta g \ll g$ (say, $\Delta g=10^{-4}g$) is chosen, and  $P(g\pm\Delta g/2)$ and $n(g)$ are evaluated, keeping $T$ and $\mu$ constant.
They are used to estimate
\eq{g20est}{
g^{(2)}(0) \approx -\frac{1}{n(g)^2}\ \left(\frac{P(g+\tfrac{1}{2}\Delta g)-P(g-\tfrac{1}{2}\Delta g)}{\Delta g}\right).
}
We found, however, that this procedure is much more unstable numerically when evaluating $S_0$ using 
\eqn{uGu} than it was for evaluating $g^{(2)}(0)$ using \eqn{g20sol}, especially at low temperatures. 

\begin{figure}
\begin{center}
\includegraphics[width=\columnwidth]{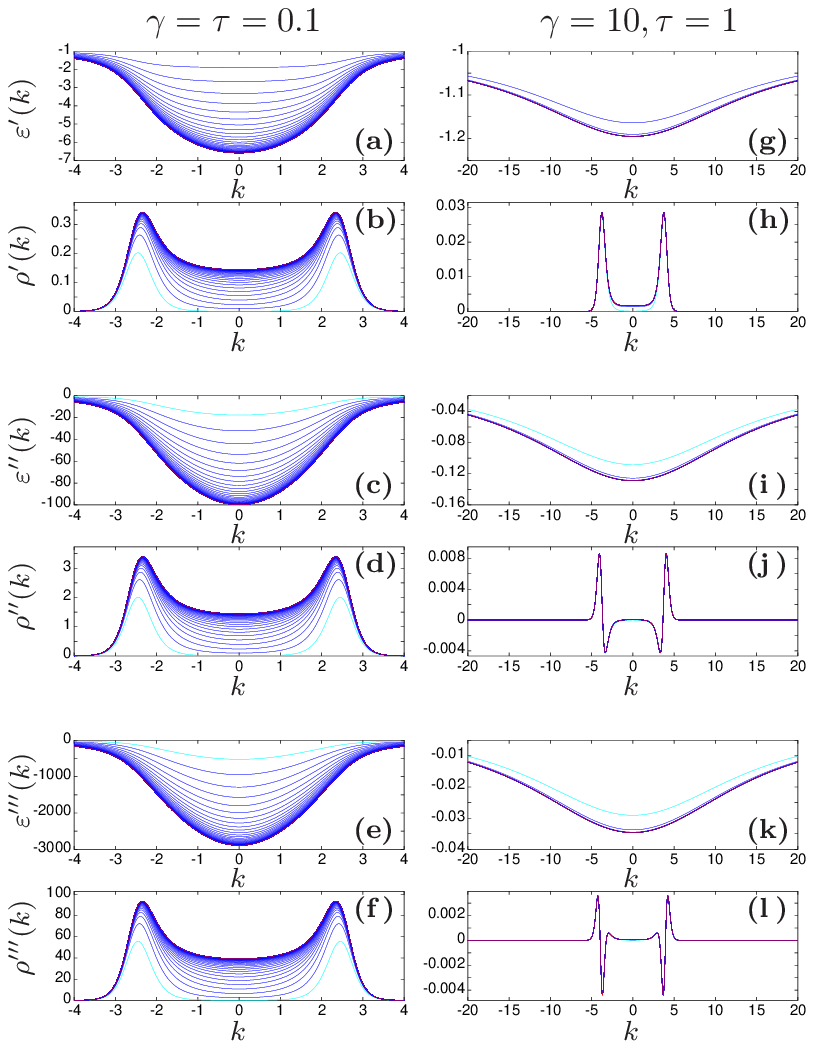}
\end{center}
\caption{
Iterations for the calculation of the derivatives $\ve'(k)$ and $\rho'(k)$ via \eqn{dek-ieq} and \eqn{drhok-ieq} (top two rows);
for the calculation of the corresponding second derivatives via \eqn{ddek-ieq} and \eqn{ddrhok-ieq} (middle two rows);
and the third derivatives via \eqn{dddek-ieq} and \eqn{dddrhok-ieq} (bottom two rows).
Notation and colors as in Fig.~\ref{fig:vars}.
The left column shows the quasicondensate case $\gamma=\tau=0.1$, while the right shows the fermionized $\gamma=10$, $\tau=1$.
}
\label{fig:iters}
\end{figure}

%%%%%%%%%%%%%%%%%%%%%%%%%%%%%%%%%%%%%%%%%%%%%%
\subsubsection{Stable approach}
A more accurate (and ultimately more efficient) approach is found as follows. 
From \eqn{nsol}, we see that 
\eq{dndu}{
\left(\frac{\partial n}{\partial\mu}\right)_{g,T} = \int_{-\infty}^{\infty} \frac{\partial\rho(k)}{\partial\mu} dk.
}
Defining 
\eq{dpdu}{
\rho'(k) = \frac{\partial\rho(k)}{\partial\mu}, 
}
and differentiating the equation \eqn{rhok-ieq} we find that $\rho'(k)$ obeys its own integral equation
\eq{drhok-ieq}{
\rho'(k)\left[1+e^{\ve(k)/T}\right] + \frac{\rho(k)\ve'(k)}{T} e^{\ve(k)/T}
= \frac{g}{\pi} \int_{-\infty}^{\infty} \frac{dq\ \rho'(q)}{g^2+(k-q)^2}.
}
where
\eq{dedu}{
\ve'(k) = \frac{\partial\ve(k)}{\partial\mu}.
}
This quantity, in turn, can be obtained from a second integral equation that comes from differentiating
\eqn{ek-ieq}: 
\eq{dek-ieq}{
\ve'(k) = -1 +\frac{g}{\pi} \int_{-\infty}^{\infty} \frac{dq\,\ve'(q)}{g^2+(k-q)^2}\ \frac{1}{1+e^{\ve(q)/T}}.
}
The above two integral equations can also be solved by iteration once we know $\ve(k)$ and $\rho(k)$.
It is most convenient to use the same numerical lattice for $k$ and $q$ in all the integral equations.
The starting forms for the iteration are
\eq{diter0}{
\ve^{\prime\,(0)}(k) = -1
}
and
\eq{diter0-rho}{
\rho^{\prime\,(0)}(k) = -\frac{1}{T}\ \frac{\rho(k)\ve'(k)}{1+e^{-\ve(k)/T}}.
}
The final result is
\eq{uGmu}{
S_0 = \frac{{\rm var} N}{N} = \frac{T}{n}\ \int_{-\infty}^{\infty} \rho'(k) dk.
}
Examples of the iterations are shown in the top two rows of Fig.~\ref{fig:iters}.

%%%%%%%%%%%%%%%%%%%%%%%%%%%%%%%%%%%%%%%%%%%%%%%%%%%%%
\subsection{Phase diagram for density grain statistics}
\label{UGRESULT}

\begin{figure}[h!]
\begin{center}
\includegraphics[width=\columnwidth]{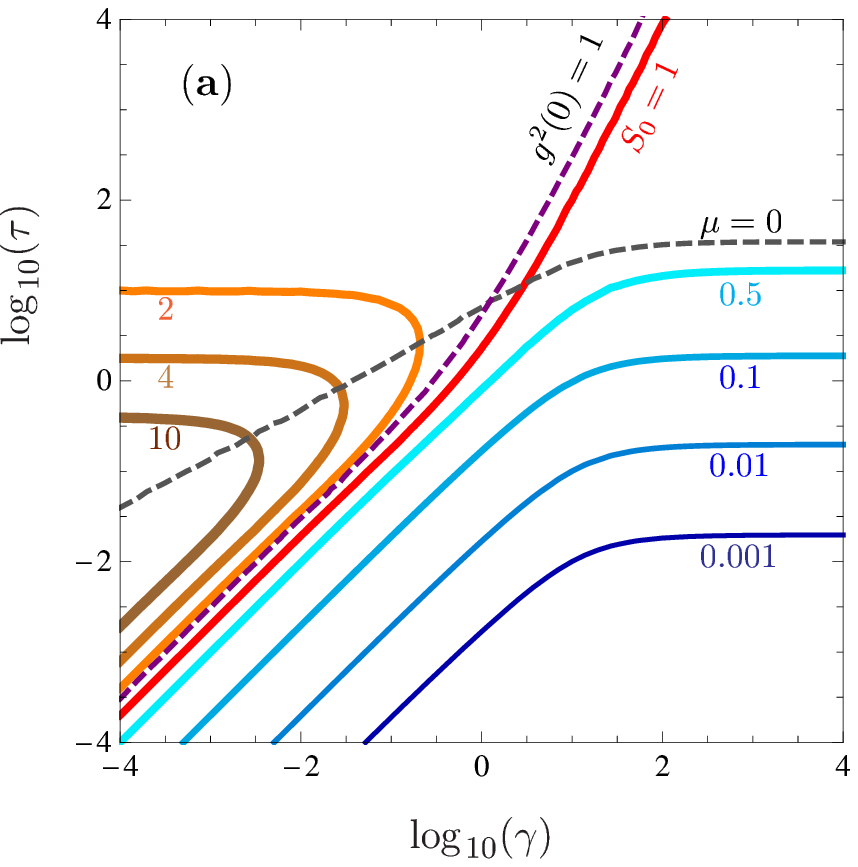}
\\
\includegraphics[width=4cm]{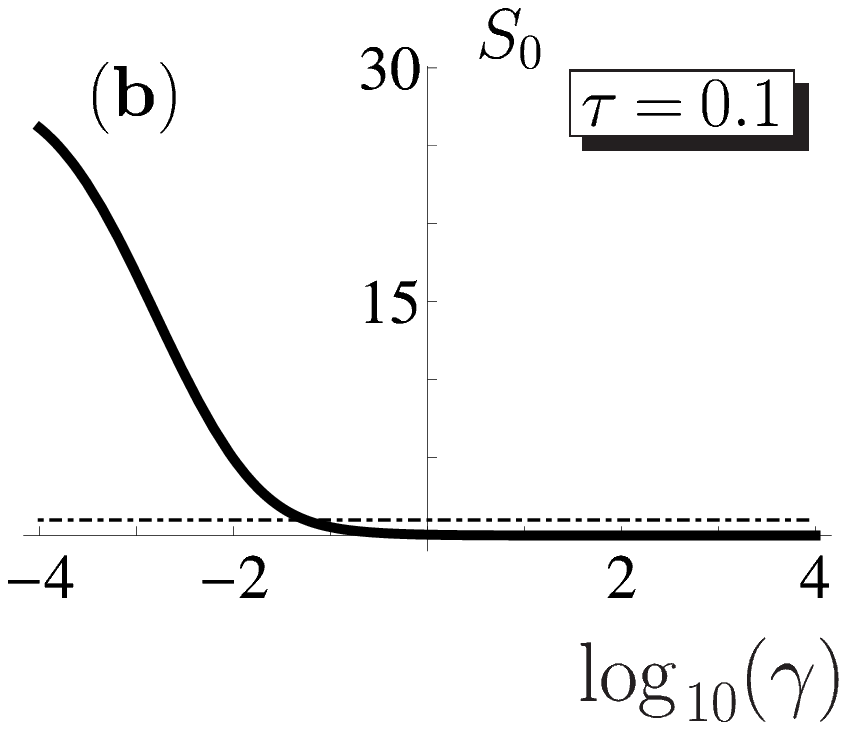}
\
\includegraphics[width=4cm]{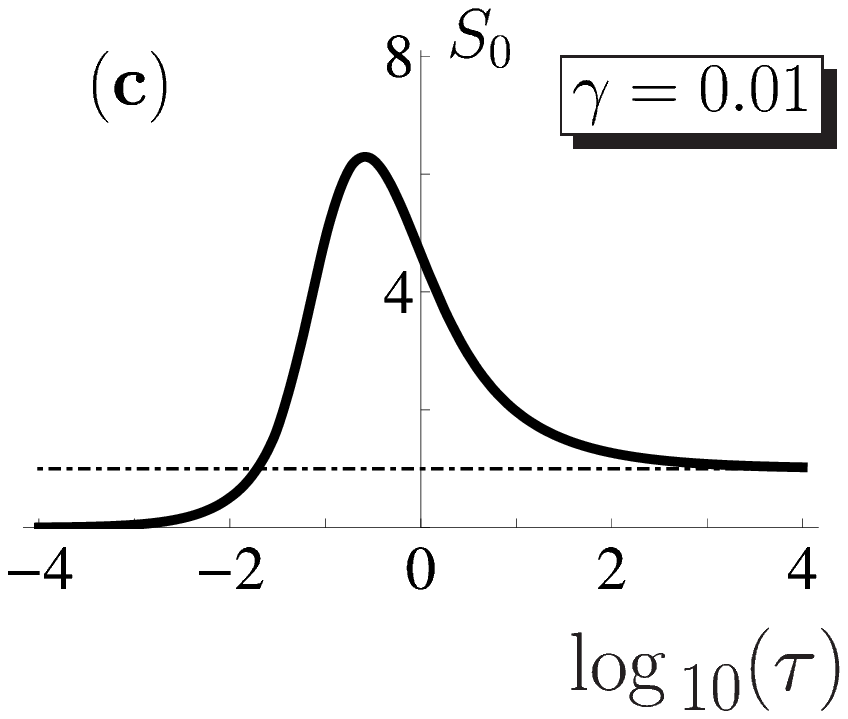}
\end{center}
\caption{
Top: Contours of $S_0$ in the parameter space of the 1d Bose gas (values of $S_0$ indicated on the plot).
This quantity indicates number variance as compared to Poissonian statistics, as well as the typical number of particles partaking 
in independent density fluctuations because $\wb{\mc{N}}\approx S_0$. Two reference lines are also shown in thick dashed style: 
The location at which perfect second order coherence occurs ($g^{(2)}(0)=1$) in purple, and the $\mu=0$ crossover in gray.
Bottom: behavior of $S_0$ along $\tau=0.1$ (left) and $\gamma=0.01$ (right). 
}
\label{fig:uG}
\end{figure}

Armed with this algorithm, we have calculated $S_0$ for a wide range of physical parameters. Its behavior is shown in 
Fig.~\ref{fig:uG} as a contour plot, along with some representative transects at constant $\gamma$ and constant $\tau$.

In the discussion below, we refer to the different physical regimes described in Sec.~\ref{GAMTAU}. 
For clarity, keep in mind that the different regimes are separated by continuous crossovers not phase transitions. All the transitions come together in the neighborhood of $\gamma=\tau=1$.

As a first point, let us note the $S_0=1$ line in Fig.~\ref{fig:uG}a that separates the regions of super-Poissonian and sub-Poissonian statistics.  
It lies at $\tau\approx2\gamma$ in the quasicondensate ($\gamma\ll1$), and shifts slope to $\tau\approx\gamma^2$ in the high-temperature, high-gamma region. The latter lies in the middle of the crossover between a normal classical gas and a high-temperature fermionized one. 
Note also that, except for a small shift, the location of $S_0=1$ follows the $g^{(2)}(0)=1$ reference line 
which separates the bunched and antibunched gases. 
Although the two properties often appear together and are correlated, they are physically distinct: 
bunching/antibunching refers to the precisely local behavior, while Poissonian or non-Poissonian statistics refers to the behavior of larger pieces of the gas. 

The \emph{shift}, is interesting, however. It the quasicondensate, $g^{(2)}(0)=1$ occurs at $\tau\approx 3\gamma$, i.e. at about a 50\% higher temperature than $S_0=1$. In the high $\gamma$ regime,
$g^{(2)}(0)=1$ occurs at $\tau\approx3\gamma^2$, i.e about 3 times higher than $S_0=1$. 
The observed shift indicates 
that the bunching/antibunching transition occurs under $S_0>1$ conditions. Due to the relationship \eqn{uGg2}, this means that $\int dz (g^{(2)}(z)-1)>0$. Since  $g^{(2)}(0)=1$ at these points, this unambiguously indicates a hump in the $g^{(2)}(z)$ function
at $z\ne0$. It also means that density-density correlations at a moderate but nonzero distance are stronger than the zero-range local ones near the transition. Such a preferred correlation length has been noted before \cite{Deuar09,Carusotto01}. The results found here show that this phenomenon
exists across the whole range of temperatures and interaction strengths in the vicinity of $g^{(2)}(0)\approx1$.
A related point to accentuate is that sub-Poissonian statistics occurs only strictly together with antibunching ($g^{(2)}(0)<1$). 
In this region quantum fluctuations are always an important contributing factor.\\

The second reference line, $\mu=0$ is also an important watershed between essentially quantum gas physics when $\mu>0$, and largely ideal-gas-like behavior when $\mu<0$.  
In the dilute gas with $\gamma\ll1$, $\mu=0$ is a dividing line
between (a) quasicondensate behavior at lower temperatures which can still be somewhat described by a generalized Bogoliubov theory \cite{Mora03} and (b) a quantum degenerate but non-quasicondensate state at higher temperatures that is better captured by a Hartree-Fock (HF) or c-field description \cite{Pietraszewicz17}. The crossover between (a) and (b) has been studied in detail by \cite{Henkel17}. In the fermionized regime, $\mu=0$ occurs at $\tau\approx40$. Certainly, for any kind of ideal-gas-like behavior $\mu<0$ is necessary, so the lower temperature $\mu>0$ region in the fermionized case is strongly nonclassical.

The most prominent feature in all three panels of Fig.~\ref{fig:uG} is the huge bulge of large $S_0\gg1$. 
On the contour plot it is shown by brownish contours and straddles the $\mu=0$ region in the $\gamma\lesssim1$ dilute gas. 
This ``bulge''  spans both the $\mu<0$ HF degenerate region and the thermal-dominated quasicondensate, including the soliton-rich and quantum turbulent regions. The statistics here is very strongly super-Poissonian. 
$S_0$ grows rapidly as the gas becomes more dilute with falling $\gamma$. We will see later that also higher moments of the density grains are very high in this regime and grow rapidly. 
Generally speaking, the two distinguishing features in the ``bulge'' are that the gas is quantum degenerate and fluctuations are dominated by thermal effects (as opposed to quantum fluctuations). 

A second remark about the ``bulge'' is that $S_0$ falls rapidly and evenly as the $g^{(2)}(0)=1$ threshold into the quantum-fluctuation-dominated quasicondensate is reached. There is no indication of plateau behavior in $S_0$, in direct contrast to the behavior of $g^{(2)}(0)$ which flattens out prominently  below $\tau\sim\sqrt{\gamma}$ \cite{Kheruntsyan03}. 

Another global feature is that both regimes 
(low-temperature quasicondensate and the fermionized gas), that are dominated by quantum fluctuations 
share the quality of being strongly sub-Poissonian. 
In the quasicondensate, one has that $\mu\approx gn$ and  $\mu/k_BT \approx 2\gamma/\tau$. 
Since  the lines of constant $S_0$ are parallel to $\gamma=\tau$, so they correspond to constant values of $\mu/k_BT$. 
In fact, inspection of the values on the contour diagram shows that 
\eq{S0q}{
S_0\approx\frac{k_BT}{\mu}
} 
is a very good match in the quasicondensate (for small and large $S_0$).
In the fermionized gas, with $\gamma\gtrsim100$, $S_0$ becomes dependent only on $\tau$ and independent of $\gamma$, and is fairly well estimated by $S_0\approx 20\tau$.

Finally, as expected, $S_0$ tends to the shot noise value of unity in the classical regime at high $\tau$, regardless of the interaction strength $\gamma$. This is best seen in the $\gamma=0.01$ transect Fig.~\ref{fig:uG}c, where the $S_0$ line passes through all three main behaviors of $S_0$.
In the high-temperature fermionized gas, the approach to $S_0$ in the classical regime is from below. We can also see that at the point when $\mu=0$, $S_0$ lies already above the value of $0.5.$

%%%%%%%%%%%%%%%%%%%%%%%%%%%%%%%%%%%%%%%%%%%%%%%%%%%%%%%%%%%%%%%%%%%%%%%%%%%%%%%%%%%%%%%%%%%%%%%%%%%%%%%%%%%%%%%%%%%%%%%%%%%%%%%%%%%%%%%%%%%%%%%%%%%%%%%%%%%%%%%%%%%%%%%%%%%%%%%%%%%%%%%%%%%%%%%%%%%%%%%%%%%%%%%%%%
\section{Absolute density grain size}
\label{GRAINS}

The mean occupation of the density fluctuation grains, $\wb{\mc{N}}$ is a matter of much physical interest. 

What we are looking for is the average number of particles partaking in the smallest independent fluctuations.
We have in mind fluctuations whose variances can be summed.  
The size of fluctuations with this property is only bounded from below, but not from above 
in general, because if one labels two independent fluctuations as a single larger one, it will also be independent. 
This is only a formal unphysical issue, though, so we are physically interested in the \emph{smallest} average occupation of independent fluctuations. 

Let us consider heuristically what is desired for a measure of the occupation of a \emph{localized} density grain.  
We are looking for a group of atoms that appear together in a single realization. It also should appear independently of other neighboring groups, so that it satisfies the independent variance requirement \eqn{N2mc2}.
Atoms that appear together in a single density grain are correlated, so a reasonable criterion for its average width should be associated with the width of the correlated feature in the $g^{(2)}(z)$ correlation function. 
This would be
\eq{w=}{
w = \frac{1}{|g^{(2)}(0)-1|}\ \int_{-\infty}^{\infty} \left[g^{(2)}(z)-1\right]\,dz.
}
The number of atoms in this region is 
then
\eq{mcN0}{
nw = \frac{S_0-1}{|g^{(2)}(0)-1|}.
}
In a bunched scenario with $g^{(2)}(0)>1$ the quantity $nw$ is a pretty good candidate for the occupation of a typical density grain.
This is the case in the hotter parts of the parameter diagram, when we haven't yet entered a quasicondensate.

However, interpretation problems arise as temperature is lowered. Around $\gamma\approx\tau$, the condensate-like case of $g^{(0)}=1$ is reached as shown in Fig.~\ref{fig:uG}. In fact, exactly at the point where we have Poissonian statistics $S_0=1$, the width $w$ of \eqn{w=} becomes undefined. This cannot mean, of course, that there are no density fluctuations (because there is shot noise), nor is the typical length of density fluctuations infinite.
In fact, the number of atoms involved in a typical fluctuation is one.

The fluctuations continue to behave differently as temperature is further lowered into the antibunched regime, where the width of the correlation function is of the order of the healing length. However, since the atoms are effectively repelled rather than attracted to each other, it is not entirely clear how many atoms there are per density fluctuation. Perhaps again a single atom?
In fact in the fermionized regime that is reached as $\gamma$ becomes large, $S_0<1$ tells us that the variance of atom number grows much slower with system size than if we had plain shot noise. 
This must mean that, on average, fluctuations are smaller than with uncorrelated atoms.

The latter observation gives an indication that when considering density fluctuation grains, it may be wiser to consider only the fluctuations away from the baseline average of $n$
to avoid unwanted effects. The amount/probability of fluctuation away from the baseline mean is quantified by the height of $g^{(2)}(z)$.
Hence, when considering only fluctuations we should apply a better normalization. 
It is warranted to use the full information contained in the magnitude of $g^{(2)}(0)-1$
rather than naively normalizing the correlation feature by $|g^{(2)}(0)-1|$ as was done in \eqn{w=}.
Taking this into account gives us the expression
\eq{nintg}{
n\int \left[g^{(2)}(z)-1\right]\,dz
}
for the number of particles in a single fluctuation. 
\eqn{nintg} can be much less than $nw$ when $g^{(2)}(0)$ is close to one -- like in the quasicondensate.
We note that such integrals of the correlation function have been found to have significant physical meaning for quantum correlations in the past \cite{Kheruntsyan12,Deuar13}.

The expression \eqn{nintg} is still unsatisfactory at low temperatures. 
It leads to a value of zero in the shot noise region, and even negative values when antibunching is present.
The point to realize here is that integrating over $g^{(2)}(z)$ can give us an estimate of the number of particles correlated \emph{with the one at $z=0$}.
However,  it does not include that one particle at the point $z=0$ that we are measuring correlations from.
Hence, to include it one should add $1$ to the average \eqn{nintg} obtained from the correlations:
\eq{heuristic}{
n\int \left[g^{(2)}(z)-1\right]\,dz + 1.
}
The equation  \eqn{heuristic} tackles the shot noise and antibunching issues at the lowest temperatures.
In the shot noise case with $g^{(2)}(z)=1$, ($\tau\approx\gamma\lesssim1$) the expression \eqn{heuristic} gives one particle per independent fluctuation, as it must. For fermionized systems ($\gamma\gg1$), \eqn{heuristic} gives values less than one, which is reasonable.
In the $\tau\to0$, $\gamma\gg1$ limit of the zero temperature Tonks-Girardeau fermionized gas, atoms are very strongly antibunched, and much more evenly distributed according to the mean density $n$, than a shot noise scenario. Only occasional small fluctuations away from this even distribution occur. 

The above heuristic arguments indicate that \eqn{heuristic} is a fairly good estimate of the average number of particles participating in a density fluctuation.
Notably, \eqn{heuristic} is nothing other than the Poissonian-discriminating quantity $S_0$ that we met in Sec.~\ref{UG}, in its correlation function expression \eqn{uGg2}.
Thus, we tentatively conclude
\eq{mcNuG}{
\wb{\mc{N}} \approx S_0.
}

The predictions of \eqn{mcNuG} turn out very reasonable when compared to physical intuitions in all regimes of the gas. 
In both the classical and condensate-like gas, shot noise is expected to be the dominant fluctuation.
 Shot noise means that not only 
is var$\mc{N}/\mc{N}=1$, but also $\mc{N}\approx 1$. 
 And, indeed, $S_0\to1$ in both cases, though for different reasons. In the classical gas, $n$ is very small rendering the value of the integral irrelevant, while in the quasicondensate,  $g^{(2)}(z)=1$ making the integrand zero.
In other regimes where $S_0$ is not unity, \eqn{mcNuG} and \eqn{N2mc3} indicate that 
\eq{varmcN2}{
{\rm var}\mc{N} \approx \wb{\mc{N}}^2 \approx S_0^2.
}
This is not suprising, because it roughly means that the width of the distribution of $\mc{N}$ is comparable to its mean value. 
This is more or less exactly what we expect for single independent fluctuations whose distribution is not affected by the central limit theorem.

Hence, $S_0$ can be interpreted in two ways:  
as an indicator of the rate of growth of $N$ as in Sec.~\ref{UG}, but also as the typical number of particles partaking in an independent density fluctuation. 
This is a second important interpretation of the plots in Fig.~\ref{fig:uG}.

With all this in mind, we can further comment on the results shown in Fig.~\ref{fig:uG}:
\begin{itemize}
\item Density grains contain many particles in the broad ``bulge''. 
This region corresponds to a degenerate gas dominated by thermal fluctuations, and much of
it contains nonperturbative fluctuations such as solitons and quantum turbulence and is well described by matter waves \cite{Pietraszewicz17}. The fluctuating lumps of density can contain very large (mesoscopic) numbers of particles. 
As the gas becomes more dilute, the number of particles per grain 
continues to grow. 
\item When varying the temperature, the  density grains with the largest number of particles occur around the $\mu=0$ crossover between quantum turbulent quasicondensate and degenerate gases. 
\item Classical shot noise with one particle per independent fluctuation rules for $\tau\gtrsim100$.
\item In the fermionized regime as well as in the lowest temperature quasicondensate, one obtains $\wb{\mc{N}}\ll1$. 
What can it mean that  there is far less than one particle per independent density fluctuation? 
\end{itemize}

Certainly, it is hard to give a good answer to the above question if particle number is
thought of as a classical quantity subject to local realism (and straightforward physical intuition). The fact that $\wb{\mc{N}}<1$ only occurs when antibunching is present may be helpful. Antibunching is inherently 
a nonclassical property which cannot be obtained e.g. using classical fields (matter waves). It is 
also an indicator that quantum fluctuations dominate the physics, and the state cannot be described classically. 

One can tentatively conjecture that such mean $\wb{\mc{N}}\ll1$ values might correspond to a superposition of the background density and a $\pm\mc{O}(1)$ particle local fluctuation that has a very small quantum amplitude. In this case the probability of observing a fluctuation at all is small, and the expectation value of the particle number fluctuation is $\ll1$. Values of $S_0\ll1$ in turn indicate var$\mc{N}\ll\mc{N}$, which is consistent with very small variation of the fluctuation amplitude around one preferred value (such as would occur with fluctuations of $\pm1$ particle). 

%%%%%%%%%%%%%%%%%%%%%%%%%%%%%%%%%%%%%%%%%%%%%%%%%%%%%%%%%%%%%%%%%%%%%%%%%%%%%%%%%%%%%%%%%%%%%%%%%%%%%%%%%%%%%%%%%%%%%%%%%%%%%%%%%%%%%%%%%%%%%%%%%%%%%%%%%%%%%%%%%%%%%%%%%%%%%%%%%%%%%%%%%%%%%%%%%%%%%%%%%%%%%%%%%%
\section{Skewness}
\label{SKEW}

%%%%%%%%%%%%%%%%%%%%%%%%%%%%%%%%%%%%%%%%%%%%%%%%%%%%%
\subsection{Thermodynamics} 
\label{THERM3}

For higher order moments, we can proceed the same way as in Sec.~\ref{UGCALC}. 
When continuing to take derivatives of \eqn{wbN2Z} for integer $a$, one finds immediately that
\eq{wbNaZ}{
\langle \op{N}^a \rangle = \frac{(k_BT)^a}{\mc{Z}}\,\left(\frac{\partial^a\mc{Z}}{\partial\mu^a}\right)_{g,T}.
}
Substituting \eqn{wbNZ}, The third order moment of fluctuations is found to be
\eq{wbN3mu}{
\langle\delta\op{N}^3\rangle = (k_BT)^2\,\left(\frac{\partial^2\langle \op{N}\rangle}{\partial\mu^2}\right)_{g,T}.
}
This can be calculated similarly to \eqn{dndu} using
\eq{dndu2}{
\left(\frac{\partial^2 n}{\partial\mu^2}\right)_{g,T} = \int_{-\infty}^{\infty} \rho''(k)\, dk
}
with the definition
\eq{dpdu2}{
\rho''(k) = \frac{\partial^2\rho(k)}{\partial\mu^2}.
}

The skewness of the distribution of $N$ is
\eq{skewness}{
s = \frac{\langle\delta\op{N}^3\rangle\ }{({\rm var}N)^{3/2}}
}
Judging by the fact that both the 3rd and 2nd moments of $\delta\op{N}$ are extensive quantities proportional to $L$, the above skewness is not an intensive thermodynamic quantity and scales as $1/\sqrt{L}$. 
It approaches zero as the system size grows.
This can be understood as an effect of the central limit theorem. Since in the thermodynamic limit we are adding many independent density grains, so it is expected that we will get a Gaussian distribution of $N$ with skewness zero. 

However, there are some  intensive thermodynamic quantities like
\eq{Th3}{
M_3 = \frac{\langle\delta N^3\rangle}{\langle\op{N}\rangle} = T^2\ \frac{\int_{-\infty}^{\infty}\ \rho''(k) dk}{\int_{-\infty}^{\infty}\ \rho(k) dk},
}
and other higher combinations like:
\eq{Th3b}{
M^{\prime}_3 = \frac{\langle\delta\op{N}^3\rangle}{{\rm var}N} 
= T\ \frac{\int_{-\infty}^{\infty}\ \rho''(k) dk}{\int_{-\infty}^{\infty}\ \rho'(k) dk},
}
that are related to the skewness. One can use them with the definition \eqn{uG} to express $s$ in the following ways:
\eq{S2}{
s = \frac{1}{\sqrt{N}}\,\frac{M'_3}{\sqrt{S_0}} = \frac{1}{\sqrt{N}}\frac{M_3}{S_0^{3/2}}.
}
Notice that $M_3/S_0^{3/2}$ describes the rate at which skewness decays with growing system size $\sqrt{N}$.

Generally it is more physically intuitive to consider the skewness of the distribution of $\mc{N}$, which is also an intensive quantity.

Proceeding like in Sec.~\ref{DGRAINS},
the 
third moment of the total atom number can be written:
\eqa{N3mc}{
\langle \op{N}^3\rangle  &=& \sum_{jj'j''}\langle\op{\mc{N}}_j\op{\mc{N}}_{j'}\op{\mc{N}}_{j''}\rangle \nonu\\
&=& p \left[\wb{\mc{N}^3} + 3(p-1)\wb{\mc{N}}\,\wb{\mc{N}^2} + (p-1)(p-2)\wb{\mc{N}}^3\right].\qquad
}
With the definition of the fluctuations in a single density grain  
\eq{delmcN}{
\delta\op{\mc{N}} = \op{\mc{N}} - \wb{\mc{N}},
}
one obtains
\eq{N3mc2}{
\langle\delta\op{N}^3\rangle = p\,\wb{\delta\mc{N}^3},
}
and afterwards, the skewness of the distribution of $\mc{N}$
\eq{mcS}{
s_{\mc{N}} = \frac{\wb{\delta\mc{N}^3}}{(\wb{\delta\mc{N}^2})^{3/2}} = s\sqrt{p} 	
= \frac{M_3}{S_0^{3/2}\sqrt{\,\wb{\mc{N}}}}. 
}
The last equality comes from substituting \eqn{S2} and \eqn{N=}.
Finally, using the expression \eqn{mcNuG}  (obtained in Sec.~\ref{GRAINS}) 
we get to the the following prediction for the density grain skewness:
\eq{mcSuG}{
s_{\mc{N}} = \frac{M_3}{S_0^2} = s_{\mc{N}}^{\rm pred}.
}

%%%%%%%%%%%%%%%%%%%%%%%%%%%%%%%%%%%%%%%%%%%%%%%%%%%%%
\subsection{Calculation}
\label{CALC3}
To evaluate \eqn{dndu2} and \eqn{wbN3mu}, we will need a new set of integral equations for the second derivatives of $\ve(k)$ and $\rho(k)$ with respect to $\mu$. These are 
\eqa{ddek-ieq}{
\lefteqn{\ve''(k) =}&\qquad\\
& \dfrac{g}{\pi} \displaystyle\int_{-\infty}^{\infty} \dfrac{dq}{g^2+(k-q)^2}
\left\{ 
\dfrac{\ve''(q)}{1+e^{\ve(q)/T}} -\dfrac{1}{T}\,\left(\dfrac{\ve'(q)}{1+e^{\ve(q)/T}}\right)^2
\right\}\nonu
}
\eqa{ddrhok-ieq}{
\left[ 2\rho'(k)\ve'(k) + \rho(k)\ve''(k) + \frac{\rho(k)\ve'(k)^2}{T}\right] \frac{e^{\ve(k)/T}}{T}+\qquad&	\nonu\\
\qquad+ \rho''(k)\left[1+e^{\ve(k)/T}\right] 
= \frac{g}{\pi} \int_{-\infty}^{\infty} \frac{dq\ \rho''(q)}{g^2+(k-q)^2}&\qquad\quad
}
so that the final expression for $\langle\delta \op{N}^3\rangle$ is 
\eq{dN3mu}{
\langle\delta \op{N}^3\rangle = T^2 L \int_{-\infty}^{\infty} \rho''(k) dk.
}
The starting iterations for numerical solution are:
\eq{dditer0}{
\ve^{\prime\prime\,(0)}(k) = -\frac{g}{\pi T} \int_{-\infty}^{\infty} \frac{dq}{g^2+(k-q)^2}\ \frac{(\ve'(q))^2}{\left(1+e^{\ve(q)/T}\right)^2}
}
\eqa{dditer0-rho}{
\lefteqn{\rho^{\prime\prime\,(0)}(k) =}&\\
& \dfrac{-1}{T\left[1+e^{-\ve(k)/T}\right]}   \left( 2\rho'(k)\ve'(k) + \rho(k)\ve''(k) + \dfrac{\rho(k)(\ve'(k))^2}{T}\right).\nonu
}
Using \eqn{dN3mu} and \eqn{uGmu} the skewnesses \eqn{S2} and \eqn{mcSuG} can be readily evaluated.

This is shown in Fig.~\ref{fig:skew}.

%%%%%%%%%%%%%%%%%%%%%%%%%%%%%%%%%%%%%%%%%%%%%%%%%%%%%
\subsection{Numerical results}
\label{SRES}

\begin{figure}
\begin{center}
\includegraphics[width=\columnwidth]{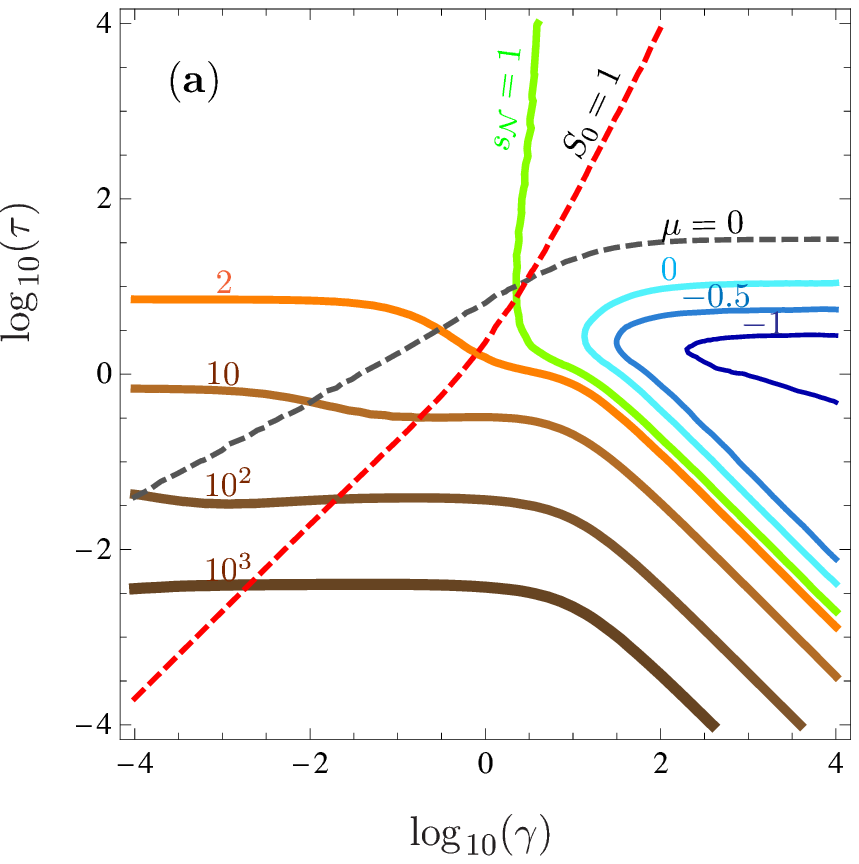}
\\
\includegraphics[width=4cm]{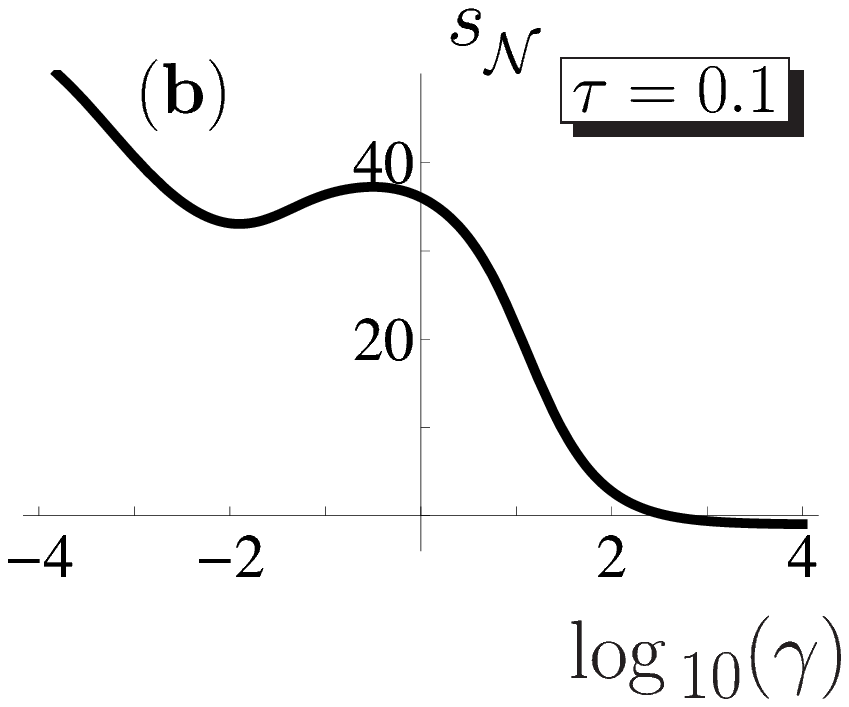}
\ 
\includegraphics[width=4cm]{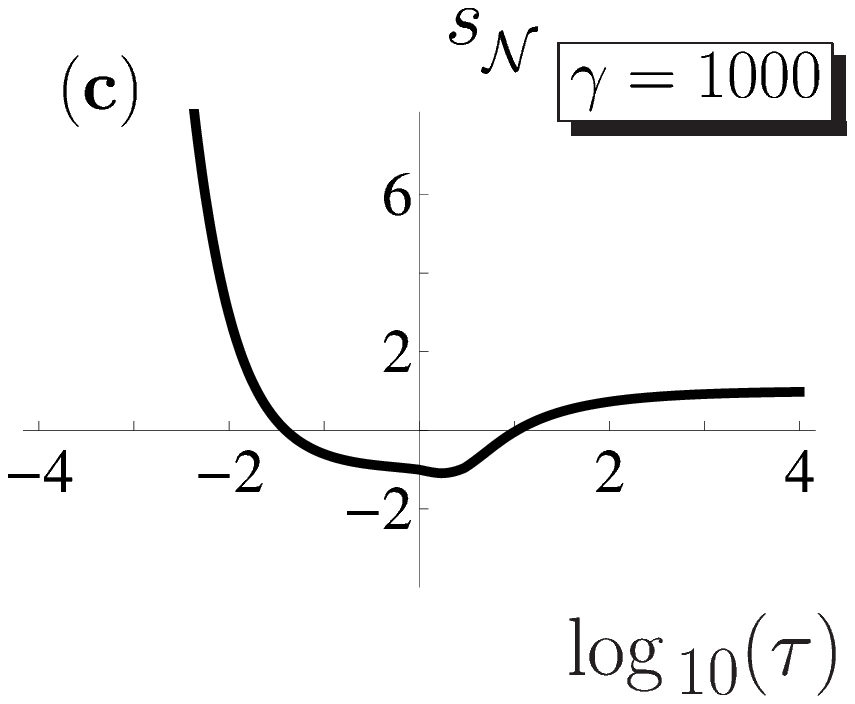}
\end{center}
\caption{
Top: Contours of the skewness of the density grain distribution $s_{\mc{N}}$ given by \eqn{mcSuG}. Values indicated on the plot. 
The location at which shot noise  occurs ($S_0=1$) is also shown in dashed red, and the $\mu=0$ crossover with dashed gray.
Bottom: behavior of $s_{\mc{N}}$ along $\tau=0.1$ (left) and $\gamma=1000$ (right).
}
\label{fig:skew}
\end{figure}

Fig.~\ref{fig:skew} shows the calculated behavior of the skewness of the density grain distribution $s_{\mc{N}}$ using \eqn{mcSuG}. 
There is also a simple correspondence between the values shown in Fig.~\ref{fig:skew} and the skewness of the total atom number $N$:  
\eq{sfig}{
s=s_{\mc{N}}^{\rm pred}\sqrt{\frac{S_0}{N}},
}
which does not invoke the heuristic arguments of Sec.~\ref{GRAINS}.
To get some bearings of known skewness values:  
an exponential distribution has a skewness of 2, while a Poisson distribution for $\mc{N}$ would have $s_{\mc{N}}=1/\sqrt{\,\wb{\mc{N}}}$.

First let us look at the region between $\tau\sim1$ and $\tau\sim\gamma$ dominated by nonperturbative thermal fluctuations, host to the $S_0\gg1$ ``bulge'' in Fig.~\ref{fig:uG}. 
The large density grains found here also have a highly (positively) skewed distribution. This indicates that not only are density grains large on average, but rare events with a much higher number of correlated particles than usual also occur. Note that since $s_{\mc{N}}\gg2$, the large $\mc{N}$ tail is much fatter than an exponential decay. It is not immediately clear what the highest likelihood/typical occupation of a density grain is in this regime, but it should be significantly less than $\wb{\mc{N}}$ when the skewness is large and positive.

While high skewness in the large $\wb{\mc{N}}$ region is perhaps not very surprising, the fact that the entire $T\to0$ limit (both quasicondensate and fermionized) has very high positive skewness is less obvious \emph{a priori}. There, it goes together with very small mean occupations $\ll1$ of the independent density fluctuations. 
We note also the presence of a nonmonotonic behavior in $s_{\mc{N}}$ around $\gamma\sim\mc{O}(1)$ which may correspond to the nonmonotonic behavior of $\mu$ noted in this vicinty recently \cite{DeRosi17}.  

A remarkable feature is the region of \emph{negative} skewness in the fermionized regime! It is interesting on several counts. 
\begin{itemize}
\item 
The negative skewness appears together with sub-Poissonian statistics.
This means that the majority of fluctuations have occupations which belong to a fairly small range at the upper end of allowed
values.
A possible explanation of the negatively skewed distribution is that the Pauli-like exclusion principle in this regime imposes
a limit on $\mc{N}$ from above. The expectation value of $\mc{N}$ is markedly less than one, which indicates that usually only
single particles take part in the fluctuations.
\item Next, some temperature-related effect must be limiting the negative skewness region to nonzero temperatures in the range $30/\gamma\lesssim\tau\lesssim10$. The cause is presently unknown.
\item That brings us to what is probably the most unexpected feature of the phase diagram in Fig.~\ref{fig:skew}a: The presence of a  crossover along $\tau\sim1/\gamma$ at all. A crossover feature in this region has not,  to our knowledge, been noted in the literature. 
Its location is around 
\eq{xover}{
\tau \approx 30/\gamma, \qquad \text{i.e.} \qquad  
k_BT \approx \frac{15\hbar^4n^3}{m^2g}.
} 
These are rather strange coefficients. So far we do not know the physics behind it. 
\end{itemize}

Finally, let us look at the classical limit of $\tau\gg1$. Here in the contour plot, skewness is seen to take small values $\mc{O}(1)$.  
In the cross-section for $\gamma=1000$ it tends to unity. In fact, $s_{\mc{N}}$ tends to unity as $\tau\gg1$ for all values of $\gamma$. This is a natural consequence of the Poissonian nature of the fluctuations in the classical gas, and their shot noise character. Since $\mc{\wb{N}}\to1$, and the distribution is Poissonian, $s_{\mc{N}}=1/\sqrt{\wb{\mc{N}}} \to 1$.

%%%%%%%%%%%%%%%%%%%%%%%%%%%%%%%%%%%%%%%%%%%%%%%%%%%%%%%%%%%%%%%%%%%%%%%%%%%%%%%%%%%%%%%%%%%%%%%%%%%%%%%%%%%%%%%%%%%%%%%%%%%%%%%%%%%%%%%%%%%%%%%%%%%%%%%%%%%%%%%%%%%%%%%%%%%%%%%%%%%%%%%%%%%%%%%%%%%%%%%%%%%%%%%%%%
\section{Kurtosis and higher moments of the distribution}
\label{KURT}

%%%%%%%%%%%%%%%%%%%%%%%%%%%%%%%%%%%%%%%%%%%%%%%%%%%%%
\subsection{Thermodynamics}
\label{THERM4}
Beyond 3rd order, the expressions start to become more complicated. 
The next relevant standardized moment is the kurtosis of the distribution of $N$: 
\eq{kurt}{
\kappa =\frac{\langle\delta\op{N}^4\rangle}{\left({\rm var} N\right)^2}.
}
It describes the relative strength of the tails, i.e. how prone the distribution is to outliers. The \emph{excess kurtosis} $(\kappa-3)$ tells
whether the tails are stronger or weaker than for a Gaussian distribution which has $\kappa=3$.

Proceeding the same way as in Sec.~\ref{THERM3}, the 4th moment of the deviation of $N$ is 
\eq{wbN4mu}{
\langle\delta \op{N}^4\rangle = (k_BT)^3\,\left(\frac{\partial^3\langle \op{N}\rangle}{\partial\mu^3}\right)_{g,T} + 3\,(k_BT)^2\left[\left(\frac{\partial\langle \op{N}\rangle}{\partial\mu}\right)_{g,T}\right]^2.
}
Evaluation of the first term requires a new higher order derivative
\eq{dndu3}{
\left(\frac{\partial^3 n}{\partial\mu^3}\right)_{g,T} = \int_{-\infty}^{\infty} \rho'''(k)\, dk,
}
with
\eq{dpdu3}{
\rho'''(k) = \frac{\partial^3\rho(k)}{\partial\mu^3},
}
whose equations appear in Sec.~\ref{CALC4}. 
A related intensive thermodynamic quantity is 
\eq{Th4}{
M_4 = \frac{(k_BT)^3}{\langle \op{N}\rangle} \left(\frac{\partial^3\langle \op{N}\rangle}{\partial\mu^3}\right)_{g,T} = (k_BT)^3\ \frac{\int_{-\infty}^{\infty}\ \rho'''(k) dk}{\int_{-\infty}^{\infty}\ \rho(k) dk}.
}

Substituting \eqn{wbN4mu} and \eqn{wbN2mu} into \eqn{kurt} one finds
\eq{kurt2}{
\kappa = 3 + k_BT\left[\frac{\partial^3\langle \op{N}\rangle}{\partial\mu^3}\right]_{g,T}\left[\frac{\partial\langle \op{N}\rangle}{\partial\mu}\right]_{g,T}^{-2}.
}
The second term goes to zero in the thermodynamic limit as $\sim1/L \sim1/N$ because each of the partial derivatives is $\propto L$. 
This is again in line with the behavior predicted by the central limit theorem for a sum of many independent contributions, since Gaussians have $\kappa=3$. 
Note that unlike at lower orders, $\frac{\langle\delta \op{N}^4\rangle}{N}$ is not now an intensive thermodynamic quantity. 

Let us study the reason of this in terms of of independent density grains as before.
The fourth moment of the total atom number can be written:
\eqa{N4mc}{
\langle \op{N}^4\rangle  &=& 
 p \Big[\wb{\mc{N}^4} + 3(p-1)\left(\wb{\mc{N}}\right)^2 + 4(p-1)\wb{\mc{N}}\,\wb{\mc{N}^3} \\
&+& 6(p-1)(p-2)\wb{\mc{N}}^2\,\wb{\mc{N}^2}+(p-1)(p-2)(p-3)\wb{\mc{N}}^4\Big].\nonu
}
Next, one obtains
\eq{N4mc2}{
\langle\delta \op{N}^4\rangle = p\,\left[\wb{\delta\mc{N}^4} -3\,\left(\wb{\delta\mc{N}^2}\right)^2\right] + 3p^2\,\left(\wb{\delta\mc{N}^2}\right)^2.
}
From this,
\eq{mcK0}{
\kappa = 3 + \frac{1}{p}\left[\kappa_{\mc{N}}-3\right]
}
which shows that 
it is the excess kurtosis of $N$ that scales as $1/p$ with the excess kurtosis of $\mc{N}$.

In terms of the intensive thermodynamic quantity $M_4$ that we can evaluate numerically,
we get
\eq{mcK}{
\kappa_{\mc{N}} = \frac{\wb{\delta\mc{N}^4}}{(\wb{\delta\mc{N}^2})^2} = 3 + \frac{M_4}{S_0^2\,\wb{\mc{N}}}. 
}
This means that $M_4/S_0^2$ describes the rate at which excess kurtosis dissipates with growing system size $N$.

\begin{figure}
\begin{center}
\includegraphics[width=\columnwidth]{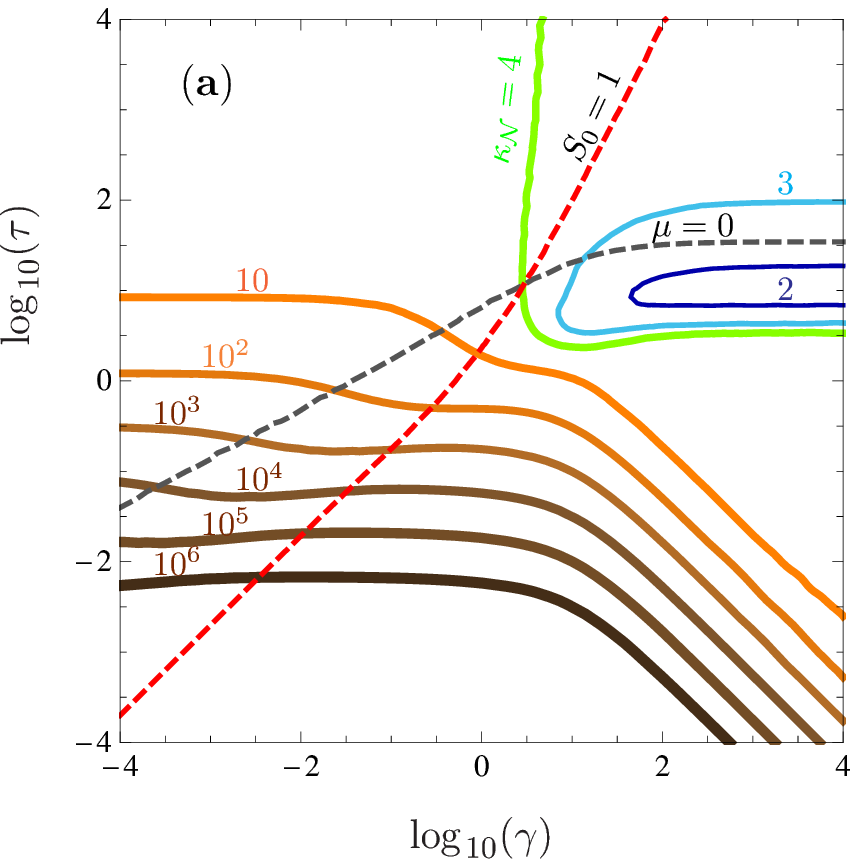}
\\
\includegraphics[width=4cm]{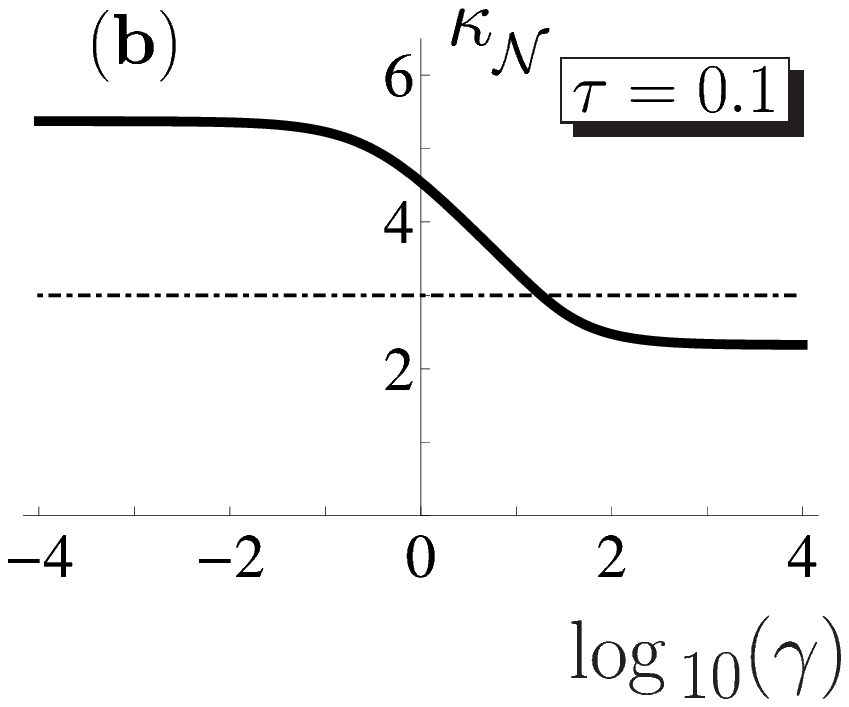}
\ 
\includegraphics[width=4cm]{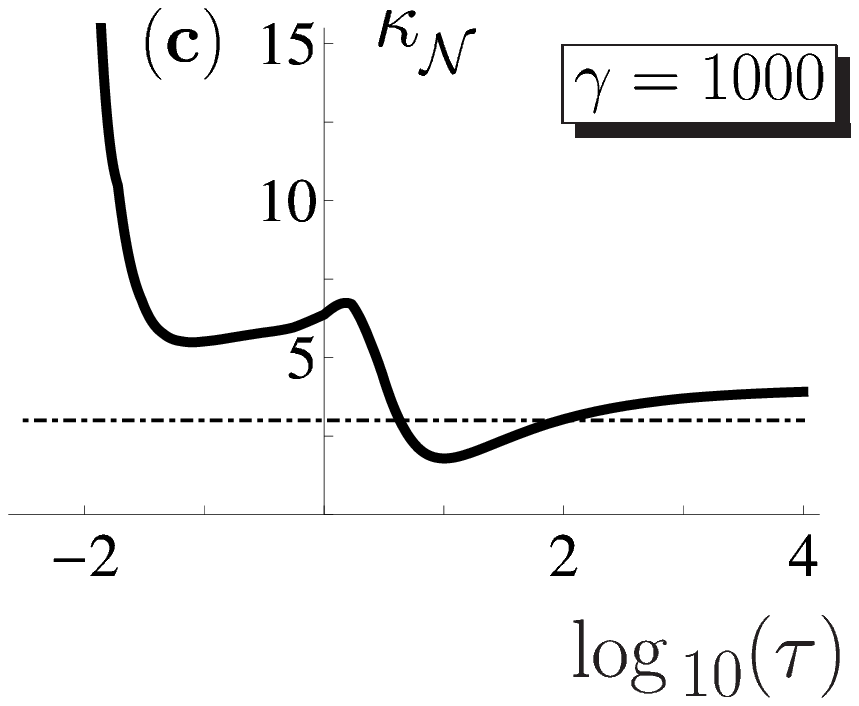}
\end{center}
\caption{
Top: Contours of the kurtosis of the density grain distribution $\kappa_{\mc{N}}$ given by \eqn{mcZuG}. Values indicated on the plot. 
The location at which shot noise  occurs ($S_0=1$) is also shown in dashed red, and the  $\mu=0$ crossover with dashed gray.
Bottom: behavior of $\kappa_{\mc{N}}$ along $\tau=30$ (left) and $\gamma=1000$ (right).
}
\label{fig:kurt}
\end{figure}

To find the kurtosis of the density grain size we use the prediction \eqn{mcNuG} for the density grain size and obtain
\eq{mcZuG}{
\kappa_{\mc{N}} = 3 + \frac{M_4}{S_0^3}=\kappa_{\mc{N}}^{\rm pred}.
}
So --- while the kurtosis of the total atom number is not an intensive thermodynamic quantity, the kurtosis of the \emph{density grains} is. As it should! 
We show its behavior in Fig.~\ref{fig:kurt}.

%%%%%%%%%%%%%%%%%%%%%%%%%%%%%%%%%%%%%%%%%%%%%%%%%%%%%
\subsection{Calculation}
\label{CALC4}
The integral equations to evaluate  \eqn{wbN4mu} and \eqn{dndu3} require another set of integral equations for the third derivatives of $\ve(k)$ and $\rho(k)$ with respect to $\mu$. 
The first is:
\eqa{dddek-ieq}{
\lefteqn{\ve'''(k) =\dfrac{g}{\pi}\displaystyle\int_{-\infty}^{\infty} \dfrac{dq}{g^2+(k-q)^2}\,\dfrac{1}{1+e^{\ve(q)/T}}\ \times}&\qquad\\
& \times
\left\{ \ve'''(q) -\dfrac{\ve''(q)\ve'(q)}{T}\,\dfrac{2+e^{\ve(q)/T}}{1+e^{\ve(q)/T}}+\dfrac{2\ve'(q)^3e^{\ve(q)/T}}{T^2(1+e^{\ve(q)/T})^2}
\right\}.\nonu
}
with starting iteration
\eqa{ddditer0}{
\ve^{\prime\prime\prime\,(0)}(k) &=& \frac{g}{\pi T} \int_{-\infty}^{\infty} \frac{dq\,\ve'(q)}{\left[(g^2+(k-q)^2\right]\left[1+e^{\ve(q)/T}\right]^2}\ \times\qquad\nonu\\ 
&&\hspace*{-1em}\times \left\{
\dfrac{2(\ve'(q))^2}{T\left(1+e^{-\ve(q)/T}\right)} -\ve''(q)\left[2+e^{\ve(q)/T}\right]
\right\}
}

The second equation is:
\eqa{dddrhok-ieq}{
&&\rho'''(k)\left[1+e^{\ve(k)/T}\right] = \frac{g}{\pi} \int_{-\infty}^{\infty} \frac{dq\ \rho'''(q)}{g^2+(k-q)^2} 
\qquad\nonu\\
&&\qquad -\ \frac{e^{\ve(k)/T}}{T}
\left\{\rho(k)\ve'''(k)+3\rho'(k)\ve''(k)+3\rho''(k)\ve'(k)\right\}\nonu\\
&&\qquad -\  \frac{3\ve'(k)e^{\ve(k)/T}}{T^2}\left[\rho(k)\ve''(k)+\rho'(k)\ve'(k)\right]\nonu\\
&&\qquad-\ \frac{\rho(k)\ve'(k)^3\,e^{\ve(k)/T}}{T^3}
}
with
\eqa{ddditer0-rho}{
\lefteqn{\rho^{\prime\prime\prime\,(0)}(k) =\dfrac{-1}{T\left[1+e^{-\ve(k)/T}\right]}\times}&\\
& \times   \Bigg\{ 
\rho(k)\ve'''(k)+3\rho'(k)\ve''(k)+3\rho''(k)\ve'(k) \nonu\\
&\qquad+ \dfrac{3\ve'(k)\left[\rho(k)\ve''(k)+\rho'(k)\ve'(k)\right]}{T} + \dfrac{\rho(k)\ve'(k)^3}{T^2}
\Bigg\}.\nonu
}

%%%%%%%%%%%%%%%%%%%%%%%%%%%%%%%%%%%%%%%%%%%%%%%%%%%%%
\subsection{Numerical results}
\label{KRES}

Fig.~\ref{fig:kurt} shows the calculated behavior of the kurtosis of the density grain distribution, $\kappa_{\mc{N}}$, using \eqn{mcZuG}. 
The kurtosis of the total atom number $N$ is related to the values shown in Fig.~\ref{fig:kurt} by
\eq{kfig}{
\kappa=3+\frac{S_0}{N}\left(\kappa^{\rm pred}_{\mc{N}}-3\right).
}
Some reference values for kurtosis are 3 for the Gaussian distribution, 6 for the exponential distribution, and $k_{\mc{N}}=3+1/\wb{\mc{N}}$ for a Poisson distribution. 
At the low end of allowable values, a square distribution has $\kappa=1.8$, whereas the lowest possible value of $\kappa=1$ is obtained with a two-peaked distribution of variable $v$ that is $P(v) = \frac{1}{2}[\delta(v-v_0)+\delta(v+v_0)]$.

Broadly speaking, the behavior of $\kappa_{\mc{N}}$ turns out to be related in a one-to one fashion to that of the skewness $s_{\mc{N}}$ --- except for the mid-temperature fermionized gas.

The very high kurtosis area is selfsame with the high skewness region in Fig.~\ref{fig:skew}. Large $\kappa_{\mc{N}}$ and  $s_{\mc{N}}$ occur together for the nonperturbative thermal region in the dilute gas, as well as the lowest temperature regimes in both the quasicondensate and the fermionized gas. The approximate relation 
\eq{skappa}{
\kappa_{\mc{N}} \approx s_{\mc{N}}^2
}
holds in this region. The unexpected crossover at \eqn{xover} is also readily visible in the kurtosis phase diagram.

In the classical gas with $\tau\gtrsim100$, $\kappa_{\mc{N}}$ tends to a value of 4 irrespective of the value  of $\gamma$. 
As with skewness, $\kappa=4$ is the expected result for a Poissonian distribution of shot noise, since $\wb{\mc{N}}\to1$ here.

Now in the fermionized gas, the triangular region that had  negative skewness does not correspond to the lowest kurtosis values. It corresponds instead to a plateau that is best seen in the $\gamma=1000$ transect of Fig.~\ref{fig:kurt}c. 
The numerical value indicates a moderate leptokurtic behavior $\kappa_{\mc{N}}\approx\mc{O}(6)$. This value, together with the negative skewness $s_{\mc{N}}$, may indicate a roughly exponential distribution. Something like  
$P(\mc{N})\sim\propto e^{\mc{N}}$ cut off at a maximum $\mc{N}_{\rm max}$, and with the tail extending towards small values. 

The most striking feature in Fig.~\ref{fig:kurt}a is the remarkable region of anomalously small kurtosis around the $\mu=0$ transition.
In this region the distribution is platykurtic ($\kappa_{\mc{N}}<3$). The lowest values of $\kappa$ do not go below the square distribution value of $\kappa_{\mc{N}}=1.8$,
but do approach closely as $\gamma\to\infty$. This behavior seems somehow intuitively reconcilable with fermionization, but we do not at this time know the reason for the platykurtic behavior.

%%%%%%%%%%%%%%%%%%%%%%%%%%%%%%%%%%%%%%%%%%%%%%%%%%%%%%%%%%%%%%%%%%%%%%%%%%%%%%%%%%%%%%%%%%%%%%%%%%%%%%%%%%%%%%%%%%%%%%%%%%%%%%%%%%%%%%%%%%%%%%%%%%%%%%%%%%%%%%%%%%%%%%%%%%%%%%%%%%%%%%%%%%%%%%%%%%%%%%%%%%%%%%%%%%
\section{Observation}
\label{OBS}

There are a number of ways that the quantities discussed here could be observed experimentally. We present three angles on this below. Moreover we believe that it would be highly interesting to see whether all the three ways of measuring the independent fluctuations do give the same self-consistent answers. This would deepen the understanding of the nature of independence of collective fluctuations in real systems.

%%%%%%%%%%%%%%%%%%%%
\subsection{Analysis of imaging bins}
\label{OBS_BIN}
One approach which is already in use is to examine the distributions and moments of the occupations of small imaging bins  
of length $\Delta$, as discussed in Sec.~\ref{UG} or \cite{Bisset13}. 
This approach amounts to looking at statistics of small bins in the gas that correspond closely to the density grains.

Some experimental groups have already made measurments  of fluctuations ${\rm var}N^{(\Delta)}$ \cite{Armijo10,Jacqmin11,Armijo11,Armijo12} and the third order moment $\langle(\delta N^{(\Delta)})^3\rangle$ \cite{Armijo10}. 
Primarily, these experiments looked at bins of width $\Delta$ that were the smallest resolvable imaging pixels, and in fact about two times smaller than the correlation length. 
To relate their results to the thermodynamic quantity $S_0$, a special empirical calibration was necessary to compensate for the correlations between adjacent bins \cite{Armijo10}. 

Calibrating for correlations should be avoidable by  using sufficiently large bins.
To do so we need to be able to use the local density approximation (LDA) $\langle|\psi(x)|^2\rangle\approx\langle\op{N}\rangle/L$ inside each bin.
Similarly to previous experiments \cite{vanAmerongen08,Armijo10,Jacqmin11}, bins 
with different densities sample the gas in different regimes of $\tau$ and $\gamma$.
To access skewness or kurtosis of the independent density grains with \eqn{mcSuG} and \eqn{mcZuG}, one should first obtain the quantities $M_3$ and $M_4$, using the same experimental images and bins as for $S_0$.

%%%%%%%%%%%%%%%%%%%%
\subsection{Analysis of finite-size scaling}
\label{OBS_FS}

Before describing this second approach, let us make some basic observations. 
The skewness of the particle number in the whole cloud and the deviation of its kurtosis from 3 are small finite-size effects:
\eq{fs3}{
s = \frac{M_3}{S_0^{3/2}\,\langle\op{N}\rangle^{1/2}}
}
\eq{fs4}{
\kappa = 3 + \frac{M_4}{S_0^2\,\langle\op{N}\rangle}
}
By studying their scaling, the properties of the local independent fluctuations can be extracted. 

At each order of fluctuation moments there is an intensive quantity which applies in the same form for both the entire system and the local independent fluctuations:
\eq{inte2}{
S_0 = \frac{\langle\delta\op{N}^2\rangle}{\langle\op{N}\rangle} = \frac{\wb{\delta\mc{N}^2}}{\wb{\mc{N}}},
}
\eq{inte3}{
M_3 = \frac{\langle\delta\op{N}^3\rangle}{\langle\op{N}\rangle} = \frac{\wb{\delta\mc{N}^3}}{\wb{\mc{N}}},
}
\eq{inte4}{
M_4 = \frac{\langle\delta\op{N}^4\rangle-3\langle\delta\op{N}^2\rangle^2}{\langle\op{N}\rangle} = \frac{\wb{\delta\mc{N}^4}-3\left(\wb{\delta\mc{N}^2}\right)^2}{\wb{\mc{N}}}.
}
Accordingly, since each of the middle expressions above has a denominator $\langle\op{N}\rangle$, 
there will be a related extensive quantity that scales proportionally to $\langle\op{N}\rangle$  
and to the size of the system $L$.

Now, consider a large segment of the cloud of length $l$. If it has practically a uniform density, then intensive quantities can be extracted from the rate at which the extensive ones change with $l$. 
Having an ensemble of density images $n_j(x)$ of successively produced clouds labeled by $j$, the way to proceed
would be as follows: 
After setting the coordinates to $x=0$ in the center of each individual imaged cloud,  
we define a family of central segments of width $l$  lying in the range $x\in [-l/2, l/2]$. Then one obtains width-dependent atom numbers for each $j$th image
\eq{Njl}{
N_j^{(l)} = \int_{-l/2}^{l/2} dx\ n_j(x).
}
Ensemble averaging gives  $\langle N^{(l)}\rangle$ and the moments $\delta N^{(l)}=N^{(l)}-\langle N^{(l)}\rangle$   
(all dependent on the segment length). 
The expressions for the quantities we are interested in take the forms:
\eq{grad2}{
S_0 = \frac{1}{n}\frac{\partial\,{\rm var}N^{(l)}}{\partial l},
}
\eq{grad3}{
M_3 = \frac{1}{n}\frac{\partial\langle(\delta N^{(l)})^3\rangle}{\partial l},
}
\eq{grad4}{
M_4 = \frac{1}{n}\frac{\partial}{\partial l}\left[\langle(\delta N^{(l)})^4\rangle-3({\rm var}N^{(l)})^2\right].
}
Now notice that $n$ is known.
By fitting the linear dependence of ${\rm var}N^{(l)}$, $\langle(\delta N^{(l)})^3\rangle$, and $\langle(\delta N^{(l)})^4\rangle$
on the analyzed box length $l$, quite robust estimates of $S_0$, $M_3$, $M_4$ can be obtained. 

Generally, one expects nonlinear dependence to emerge at the low end of $l$ (due to local density correlations) and at the high end of $l$ (due to non-uniformity of the ensemble-averaged density). Such marginal values of $l$ would have to be ignored in the analysis.
The key  advantage of using the gradients \eqn{grad2}-\eqn{grad4} over direct evaluation of \eqn{inte2}-\eqn{inte4} is that one can readily inspect whether indeed the linear extensive regime is present.

Once we have the fitted estimates for $S_0$, $M_3$, and $M_4$, properties of the density grain distribution can be obtained  using  \eqn{mcSuG} and \eqn{mcZuG}.

%%%%%%%%%%%%%%%%%%%%
\subsection{Analysis of single shot images}
\label{OBS_SHOT}
A third and arguably most interesting angle for an experimental proposal is 
to determine to what degree the intuitive picture of localized independent fluctuations leading to \eqn{mcNuG} is in fact accurate. 
If it is true, then independent fluctuations of widths comparable to the $g^{(2)}(z)$ correlation function should be visible at the level of individual cloud images. 
It might not be true if the arguments for \eqn{mcNuG} break down. For example because the spatial extent of an independent number fluctuation is larger than the separation between their centers of mass. Which of these cases occurs may depend on the physical parameters $\gamma$ and $\tau$.

Sec.~\ref{OBS_BIN} considered single-shot variations in arbitrarily chosen bins with artificially sharp boundaries. However, in the intuitive picture one expects that the independent localized fluctuations should be visible in images as density concentrations (``lumps'', with varying but naturally-set widths).
These lumps should be discernible and countable without the need to impose artificially sharp bins.
Confirming or refuting the accuracy of such a \emph{single-shot lump} interpretation would help in understanding how to interpret independence at the level of physical intuition. 

The issue of  how well such counting of lumps corresponds to the  independent domains in a gas has been a long-standing point of discussion in the field.
The same concerns the width of the density-density correlation function $g^{(2)}(z)$, and how well it corresponds to the widths of such lumps.  
One way to go about checking the accuracy of these conjectures is to compare local density to the background mean density. 
Take the following definition of a background density for each $j$th experimental shot: 
\eq{nj}{
\wb{n}_j = \frac{1}{l}\int_{-l/2}^{l/2} dx\, n_j(x),
}
assuming that a central piece of the cloud of length $l$ is  reasonably long. 
The averaging in \eqn{nj}
is spatial and independent of the full ensemble average.
Then, the lump boundaries $x_{(p)}$ can be set  by
$n_j(x_{(p)})=\wb{n}_j$ 
at the crossing points between the local density and the mean one. 
The occupation of the $p$th lump would be taken as the number of atoms between consecutive lump boundaries, i.e.
\eq{njp}{
N_{jp} = \int_{x_{(p)}}^{x_{(p+1)}}dx\ n_j(x).
}
Adjacent ``lumps'' would consist of alternately above-average-density and below-average-density sections,
but adjacent $N_{jp}$ would not  necessarily take on alternating high and low values because the lump widths can differ.  
Note that with this whole procedure, the pre-set bin positions are avoided.
If the intuitive localized density fluctuation picture and the reasoning for \eqn{mcNuG} is correct then the mean, variance, skewness and kurtosis of the lump occupations $N_{jp}$ 
should agree with $\wb{\mc{N}}$, ${\rm var}\mc{N}$, $s_{\mc{N}}$, and $\kappa_{\mc{N}}$ obtained using the methods in the previous subsections~\ref{OBS_BIN} and~\ref{OBS_FS}.

Some practical points to consider for such a test include:
\begin{enumerate}
\item The need to have sufficient resolution to resolve adjacent density concentrations, such as in the Palaiseau experiments \cite{Armijo10,Jacqmin11,Armijo11,Jacqmin11,Armijo12,Jacqmin12}.
\item It is probably necessary to work in a regime with $\wb{\mc{N}}\gg1$ such as the quasicondensate, so that all or most relevant fluctuations lead to an actual imaging signal.
\item A sufficiently wide cloud is required to catch a sizable number of lumps in the uniform central section.
\end{enumerate}

%%%%%%%%%%%%%%%%%%%%%%%%%%%%%%%%%%%%%%%%%%%%%%%%%%%%%%%%%%%%%%%%%%%%%%%%%%%%%%%%%%%%%%%%%%%%%%%%%%%%%%%%%%%%%%%%%%%%%%%%%%%%%%%%%%%%%%%%%%%%%%%%%%%%%%%%%%%%%%%%%%%%%%%%%%%%%%%%%%%%%%%%%%%%%%%%%%%%%%%%%%%%%%%%%%
\section{Conclusions}
\label{CONCLUSIONS}

Fresh exact quantum results for the 1d interacting gas are rather rare to come by.
Generally, the known findings concern global quantities or two-body correlations. 
The distinguishing feature of this work is that it provides exact results for a different kind of quantity that describes the mesoscale, and goes beyond two-body correlations. 

We started with results for moments of the total atom number. Calculations were made using the additional integral equations \eqn{drhok-ieq}, \eqn{dek-ieq}, \eqn{uGmu} at 2nd order, and those in Secs.~\ref{CALC3} and~\ref{CALC4} for 3rd and 4th. However, by then taking advantage of the definition of independence (that independent contributions to the global variance add), we arrived at quantities that describe the properties of individual mesoscopic density grains.

Our main physical results are summarized by three figures that present a phase diagram for the behavior of the independent density grains. Fig.~\ref{fig:uG} shows the mean particle number $\wb{\mc{N}}$ in an independent density fluctuation, and the super/sub-Poissonian nature of the distribution $P(\mc{N})$. 
Figs.~\ref{fig:skew} and~\ref{fig:kurt} show the main remaining qualities of $P(\mc{N})$: its skewness and kurtosis. Neither of these can be obtained from studying two-body correlations.
The figures also give direct access to the moments of the total atom number $N$ through the relations ${\rm var}N=N S_0$, \eqn{sfig} and \eqn{kfig},
 without reference to the arguments made for \eqn{mcNuG}.

Overall, one can distinguish several regions: 
\begin{enumerate}
\item A region with large ``mesoscopic'' density grains in which  many particles take part in a single density fluctuation. 
This includes the quantum turbulent regime, the thermal-dominated quasicondensate, and the degenerate incoherent regime with $\mu<0$ well described by Hartree-Fock theory.
The distribution of $\mc{N}$ has long positive tails here, such that some fluctuations have extra large occupations.
\item The lowest temperatures host sub-Poissonian quantum fluctuations with $\wb{\mc{N}}\ll1$ and obey a distribution with high skewness and kurtosis. One can observe similar behavior in the dilute bosonic gas and the strongly fermionized regime.
\item The mid-temperature regime of the fermionized gas is characterized by a negatively skewed and moderately leptokurtic distribution.
\item A platykurtic regime in the fermionized gas near the region where $\mu=0$.
\item The classical region at large $\tau$, displaying single-particle shot noise. 
\end{enumerate}

The above list includes some phenomena whose in-depth causes are unexplained for now. This concerns  the exact mechanisms behind the low $\mc{N}$ in the mid-temperature fermionized gas as well as the negative skewness and platykurtic behaviors seen there. Moreover, also the physical cause of the unexpected crossover around $\tau\sim1/\gamma$ in the fermionized gas is not obvious.

The method presented in this paper is readily extensible to higher order moments, although beyond kurtosis the expressions get more and more complicated, while the physical interpretation less and less intuitive. A really interesting question would be whether a similar approach could be found to study the \emph{phase} domains in the gas.

At the end in Sec.~\ref{OBS}, we have outlined several ways to measure the properties of the density grains experimentally.  
A better understanding of the relationship between single shot manifestations of independent fluctuations and ensemble averaged ones would be valuable conceptually and in practice.
We have proposed a way to study this quantitatively in Sec.~\ref{OBS_SHOT}.

%%%%%%%%%%%%%%%%%%%%%%%%%%%%%%%%%%%%%%%%%%%%%%%%%%%%%%%%%%%%%%%%%%%%%
%Acknowledgments
%%%%%%%%%%%%%%%%%%%%%%%%%%%%%%%%%%%%%%%%%%%%%%%%%%%%%%%%%%%%%%%%%%%%%
\acknowledgments
We are grateful to Karen Kheruntsyan for discussions and insights on the Yang-Yang exact solutions, and for sharing the baseline program we used to self-consistently solve the integral equations before introducing refinements.
 This work was supported by the National Science Centre grant No. 2012/07/E/ST2/01389.

\bibliography{cfields}

\begin{thebibliography}{55}%
\makeatletter
\providecommand \@ifxundefined [1]{%
 \@ifx{#1\undefined}
}%
\providecommand \@ifnum [1]{%
 \ifnum #1\expandafter \@firstoftwo
 \else \expandafter \@secondoftwo
 \fi
}%
\providecommand \@ifx [1]{%
 \ifx #1\expandafter \@firstoftwo
 \else \expandafter \@secondoftwo
 \fi
}%
\providecommand \natexlab [1]{#1}%
\providecommand \enquote  [1]{``#1''}%
\providecommand \bibnamefont  [1]{#1}%
\providecommand \bibfnamefont [1]{#1}%
\providecommand \citenamefont [1]{#1}%
\providecommand \href@noop [0]{\@secondoftwo}%
\providecommand \href [0]{\begingroup \@sanitize@url \@href}%
\providecommand \@href[1]{\@@startlink{#1}\@@href}%
\providecommand \@@href[1]{\endgroup#1\@@endlink}%
\providecommand \@sanitize@url [0]{\catcode `\\12\catcode `\$12\catcode
  `\&12\catcode `\#12\catcode `\^12\catcode `\_12\catcode `\%12\relax}%
\providecommand \@@startlink[1]{}%
\providecommand \@@endlink[0]{}%
\providecommand \url  [0]{\begingroup\@sanitize@url \@url }%
\providecommand \@url [1]{\endgroup\@href {#1}{\urlprefix }}%
\providecommand \urlprefix  [0]{URL }%
\providecommand \Eprint [0]{\href }%
\providecommand \doibase [0]{http://dx.doi.org/}%
\providecommand \selectlanguage [0]{\@gobble}%
\providecommand \bibinfo  [0]{\@secondoftwo}%
\providecommand \bibfield  [0]{\@secondoftwo}%
\providecommand \translation [1]{[#1]}%
\providecommand \BibitemOpen [0]{}%
\providecommand \bibitemStop [0]{}%
\providecommand \bibitemNoStop [0]{.\EOS\space}%
\providecommand \EOS [0]{\spacefactor3000\relax}%
\providecommand \BibitemShut  [1]{\csname bibitem#1\endcsname}%
\let\auto@bib@innerbib\@empty
%</preamble>
\bibitem [{\citenamefont {Sadler}\ \emph {et~al.}(2006)\citenamefont {Sadler},
  \citenamefont {Higbie}, \citenamefont {Leslie}, \citenamefont
  {Vengalattore},\ and\ \citenamefont {Stamper-Kurn}}]{Sadler06}%
  \BibitemOpen
  \bibfield  {author} {\bibinfo {author} {\bibfnamefont {L.~E.}\ \bibnamefont
  {Sadler}}, \bibinfo {author} {\bibfnamefont {J.~M.}\ \bibnamefont {Higbie}},
  \bibinfo {author} {\bibfnamefont {S.~R.}\ \bibnamefont {Leslie}}, \bibinfo
  {author} {\bibfnamefont {M.}~\bibnamefont {Vengalattore}}, \ and\ \bibinfo
  {author} {\bibfnamefont {D.~M.}\ \bibnamefont {Stamper-Kurn}},\ }\href
  {\doibase {10.1038/nature05094}} {\bibfield  {journal} {\bibinfo  {journal}
  {Nature}\ }\textbf {\bibinfo {volume} {443}},\ \bibinfo {pages} {312}
  (\bibinfo {year} {2006})}\BibitemShut {NoStop}%
\bibitem [{\citenamefont {Vinit}\ \emph {et~al.}(2013)\citenamefont {Vinit},
  \citenamefont {Bookjans}, \citenamefont {S\'a~de Melo},\ and\ \citenamefont
  {Raman}}]{Vinit13}%
  \BibitemOpen
  \bibfield  {author} {\bibinfo {author} {\bibfnamefont {A.}~\bibnamefont
  {Vinit}}, \bibinfo {author} {\bibfnamefont {E.~M.}\ \bibnamefont {Bookjans}},
  \bibinfo {author} {\bibfnamefont {C.~A.~R.}\ \bibnamefont {S\'a~de Melo}}, \
  and\ \bibinfo {author} {\bibfnamefont {C.}~\bibnamefont {Raman}},\ }\href
  {\doibase 10.1103/PhysRevLett.110.165301} {\bibfield  {journal} {\bibinfo
  {journal} {Phys. Rev. Lett.}\ }\textbf {\bibinfo {volume} {110}},\ \bibinfo
  {pages} {165301} (\bibinfo {year} {2013})}\BibitemShut {NoStop}%
\bibitem [{\citenamefont {De}\ \emph {et~al.}(2014)\citenamefont {De},
  \citenamefont {Campbell}, \citenamefont {Price}, \citenamefont {Putra},
  \citenamefont {Anderson},\ and\ \citenamefont {Spielman}}]{De14}%
  \BibitemOpen
  \bibfield  {author} {\bibinfo {author} {\bibfnamefont {S.}~\bibnamefont
  {De}}, \bibinfo {author} {\bibfnamefont {D.~L.}\ \bibnamefont {Campbell}},
  \bibinfo {author} {\bibfnamefont {R.~M.}\ \bibnamefont {Price}}, \bibinfo
  {author} {\bibfnamefont {A.}~\bibnamefont {Putra}}, \bibinfo {author}
  {\bibfnamefont {B.~M.}\ \bibnamefont {Anderson}}, \ and\ \bibinfo {author}
  {\bibfnamefont {I.~B.}\ \bibnamefont {Spielman}},\ }\href {\doibase
  10.1103/PhysRevA.89.033631} {\bibfield  {journal} {\bibinfo  {journal} {Phys.
  Rev. A}\ }\textbf {\bibinfo {volume} {89}},\ \bibinfo {pages} {033631}
  (\bibinfo {year} {2014})}\BibitemShut {NoStop}%
\bibitem [{\citenamefont {Petrov}\ \emph {et~al.}(2000)\citenamefont {Petrov},
  \citenamefont {Shlyapnikov},\ and\ \citenamefont {Walraven}}]{Petrov00}%
  \BibitemOpen
  \bibfield  {author} {\bibinfo {author} {\bibfnamefont {D.~S.}\ \bibnamefont
  {Petrov}}, \bibinfo {author} {\bibfnamefont {G.~V.}\ \bibnamefont
  {Shlyapnikov}}, \ and\ \bibinfo {author} {\bibfnamefont {J.~T.~M.}\
  \bibnamefont {Walraven}},\ }\href {\doibase 10.1103/PhysRevLett.85.3745}
  {\bibfield  {journal} {\bibinfo  {journal} {Phys. Rev. Lett.}\ }\textbf
  {\bibinfo {volume} {85}},\ \bibinfo {pages} {3745} (\bibinfo {year}
  {2000})}\BibitemShut {NoStop}%
\bibitem [{\citenamefont {Dettmer}\ \emph {et~al.}(2001)\citenamefont
  {Dettmer}, \citenamefont {Hellweg}, \citenamefont {Ryytty}, \citenamefont
  {Arlt}, \citenamefont {Ertmer}, \citenamefont {Sengstock}, \citenamefont
  {Petrov}, \citenamefont {Shlyapnikov}, \citenamefont {Kreutzmann},
  \citenamefont {Santos},\ and\ \citenamefont {Lewenstein}}]{Dettmer01}%
  \BibitemOpen
  \bibfield  {author} {\bibinfo {author} {\bibfnamefont {S.}~\bibnamefont
  {Dettmer}}, \bibinfo {author} {\bibfnamefont {D.}~\bibnamefont {Hellweg}},
  \bibinfo {author} {\bibfnamefont {P.}~\bibnamefont {Ryytty}}, \bibinfo
  {author} {\bibfnamefont {J.~J.}\ \bibnamefont {Arlt}}, \bibinfo {author}
  {\bibfnamefont {W.}~\bibnamefont {Ertmer}}, \bibinfo {author} {\bibfnamefont
  {K.}~\bibnamefont {Sengstock}}, \bibinfo {author} {\bibfnamefont {D.~S.}\
  \bibnamefont {Petrov}}, \bibinfo {author} {\bibfnamefont {G.~V.}\
  \bibnamefont {Shlyapnikov}}, \bibinfo {author} {\bibfnamefont
  {H.}~\bibnamefont {Kreutzmann}}, \bibinfo {author} {\bibfnamefont
  {L.}~\bibnamefont {Santos}}, \ and\ \bibinfo {author} {\bibfnamefont
  {M.}~\bibnamefont {Lewenstein}},\ }\href {\doibase
  10.1103/PhysRevLett.87.160406} {\bibfield  {journal} {\bibinfo  {journal}
  {Phys. Rev. Lett.}\ }\textbf {\bibinfo {volume} {87}},\ \bibinfo {pages}
  {160406} (\bibinfo {year} {2001})}\BibitemShut {NoStop}%
\bibitem [{\citenamefont {Hellweg}\ \emph {et~al.}(2003)\citenamefont
  {Hellweg}, \citenamefont {Cacciapuoti}, \citenamefont {Kottke}, \citenamefont
  {Schulte}, \citenamefont {Sengstock}, \citenamefont {Ertmer},\ and\
  \citenamefont {Arlt}}]{Hellweg03}%
  \BibitemOpen
  \bibfield  {author} {\bibinfo {author} {\bibfnamefont {D.}~\bibnamefont
  {Hellweg}}, \bibinfo {author} {\bibfnamefont {L.}~\bibnamefont
  {Cacciapuoti}}, \bibinfo {author} {\bibfnamefont {M.}~\bibnamefont {Kottke}},
  \bibinfo {author} {\bibfnamefont {T.}~\bibnamefont {Schulte}}, \bibinfo
  {author} {\bibfnamefont {K.}~\bibnamefont {Sengstock}}, \bibinfo {author}
  {\bibfnamefont {W.}~\bibnamefont {Ertmer}}, \ and\ \bibinfo {author}
  {\bibfnamefont {J.~J.}\ \bibnamefont {Arlt}},\ }\href {\doibase
  10.1103/PhysRevLett.91.010406} {\bibfield  {journal} {\bibinfo  {journal}
  {Phys. Rev. Lett.}\ }\textbf {\bibinfo {volume} {91}},\ \bibinfo {pages}
  {010406} (\bibinfo {year} {2003})}\BibitemShut {NoStop}%
\bibitem [{\citenamefont {Gring}\ \emph {et~al.}(2012)\citenamefont {Gring},
  \citenamefont {Kuhnert}, \citenamefont {Langen}, \citenamefont {Kitagawa},
  \citenamefont {Rauer}, \citenamefont {Schreitl}, \citenamefont {Mazets},
  \citenamefont {Smith}, \citenamefont {Demler},\ and\ \citenamefont
  {Schmiedmayer}}]{Gring12}%
  \BibitemOpen
  \bibfield  {author} {\bibinfo {author} {\bibfnamefont {M.}~\bibnamefont
  {Gring}}, \bibinfo {author} {\bibfnamefont {M.}~\bibnamefont {Kuhnert}},
  \bibinfo {author} {\bibfnamefont {T.}~\bibnamefont {Langen}}, \bibinfo
  {author} {\bibfnamefont {T.}~\bibnamefont {Kitagawa}}, \bibinfo {author}
  {\bibfnamefont {B.}~\bibnamefont {Rauer}}, \bibinfo {author} {\bibfnamefont
  {M.}~\bibnamefont {Schreitl}}, \bibinfo {author} {\bibfnamefont
  {I.}~\bibnamefont {Mazets}}, \bibinfo {author} {\bibfnamefont {D.~A.}\
  \bibnamefont {Smith}}, \bibinfo {author} {\bibfnamefont {E.}~\bibnamefont
  {Demler}}, \ and\ \bibinfo {author} {\bibfnamefont {J.}~\bibnamefont
  {Schmiedmayer}},\ }\href {\doibase 10.1126/science.1224953} {\bibfield
  {journal} {\bibinfo  {journal} {Science}\ }\textbf {\bibinfo {volume}
  {337}},\ \bibinfo {pages} {1318} (\bibinfo {year} {2012})}\BibitemShut
  {NoStop}%
\bibitem [{\citenamefont {Deuar}(2016)}]{Deuar16}%
  \BibitemOpen
  \bibfield  {author} {\bibinfo {author} {\bibfnamefont {P.}~\bibnamefont
  {Deuar}},\ }\href {\doibase http://dx.doi.org/10.1016/j.cpc.2016.08.004}
  {\bibfield  {journal} {\bibinfo  {journal} {Computer Physics Communications}\
  }\textbf {\bibinfo {volume} {208}},\ \bibinfo {pages} {92 } (\bibinfo {year}
  {2016})}\BibitemShut {NoStop}%
\bibitem [{\citenamefont {Karpiuk}\ \emph {et~al.}(2012)\citenamefont
  {Karpiuk}, \citenamefont {Deuar}, \citenamefont {Bienias}, \citenamefont
  {Witkowska}, \citenamefont {Paw\l{}owski}, \citenamefont {Gajda},
  \citenamefont {Rz\k{a}\.{z}ewski},\ and\ \citenamefont
  {Brewczyk}}]{Karpiuk12}%
  \BibitemOpen
  \bibfield  {author} {\bibinfo {author} {\bibfnamefont {T.}~\bibnamefont
  {Karpiuk}}, \bibinfo {author} {\bibfnamefont {P.}~\bibnamefont {Deuar}},
  \bibinfo {author} {\bibfnamefont {P.}~\bibnamefont {Bienias}}, \bibinfo
  {author} {\bibfnamefont {E.}~\bibnamefont {Witkowska}}, \bibinfo {author}
  {\bibfnamefont {K.}~\bibnamefont {Paw\l{}owski}}, \bibinfo {author}
  {\bibfnamefont {M.}~\bibnamefont {Gajda}}, \bibinfo {author} {\bibfnamefont
  {K.}~\bibnamefont {Rz\k{a}\.{z}ewski}}, \ and\ \bibinfo {author}
  {\bibfnamefont {M.}~\bibnamefont {Brewczyk}},\ }\href {\doibase
  10.1103/PhysRevLett.109.205302} {\bibfield  {journal} {\bibinfo  {journal}
  {Phys. Rev. Lett.}\ }\textbf {\bibinfo {volume} {109}},\ \bibinfo {pages}
  {205302} (\bibinfo {year} {2012})}\BibitemShut {NoStop}%
\bibitem [{\citenamefont {Karpiuk}\ \emph {et~al.}(2015)\citenamefont
  {Karpiuk}, \citenamefont {Sowi\ifmmode~\acute{n}\else \'{n}\fi{}ski},
  \citenamefont {Gajda}, \citenamefont {Rz\k{a}\ifmmode~\dot{z}\else
  \.{z}\fi{}ewski},\ and\ \citenamefont {Brewczyk}}]{Karpiuk15}%
  \BibitemOpen
  \bibfield  {author} {\bibinfo {author} {\bibfnamefont {T.}~\bibnamefont
  {Karpiuk}}, \bibinfo {author} {\bibfnamefont {T.}~\bibnamefont
  {Sowi\ifmmode~\acute{n}\else \'{n}\fi{}ski}}, \bibinfo {author}
  {\bibfnamefont {M.}~\bibnamefont {Gajda}}, \bibinfo {author} {\bibfnamefont
  {K.}~\bibnamefont {Rz\k{a}\ifmmode~\dot{z}\else \.{z}\fi{}ewski}}, \ and\
  \bibinfo {author} {\bibfnamefont {M.}~\bibnamefont {Brewczyk}},\ }\href
  {\doibase 10.1103/PhysRevA.91.013621} {\bibfield  {journal} {\bibinfo
  {journal} {Phys. Rev. A}\ }\textbf {\bibinfo {volume} {91}},\ \bibinfo
  {pages} {013621} (\bibinfo {year} {2015})}\BibitemShut {NoStop}%
\bibitem [{\citenamefont {Gawryluk}\ \emph {et~al.}(2017)\citenamefont
  {Gawryluk}, \citenamefont {Brewczyk},\ and\ \citenamefont
  {Rz\k{a}\ifmmode~\dot{z}\else \.{z}\fi{}ewski}}]{Gawryluk17}%
  \BibitemOpen
  \bibfield  {author} {\bibinfo {author} {\bibfnamefont {K.}~\bibnamefont
  {Gawryluk}}, \bibinfo {author} {\bibfnamefont {M.}~\bibnamefont {Brewczyk}},
  \ and\ \bibinfo {author} {\bibfnamefont {K.}~\bibnamefont
  {Rz\k{a}\ifmmode~\dot{z}\else \.{z}\fi{}ewski}},\ }\href {\doibase
  10.1103/PhysRevA.95.043612} {\bibfield  {journal} {\bibinfo  {journal} {Phys.
  Rev. A}\ }\textbf {\bibinfo {volume} {95}},\ \bibinfo {pages} {043612}
  (\bibinfo {year} {2017})}\BibitemShut {NoStop}%
\bibitem [{\citenamefont {Nowicki}\ \emph {et~al.}(2017)\citenamefont
  {Nowicki}, \citenamefont {Pietraszewicz},\ and\ \citenamefont
  {Deuar}}]{Nowicki17}%
  \BibitemOpen
  \bibfield  {author} {\bibinfo {author} {\bibfnamefont {I.}~\bibnamefont
  {Nowicki}}, \bibinfo {author} {\bibfnamefont {J.}~\bibnamefont
  {Pietraszewicz}}, \ and\ \bibinfo {author} {\bibfnamefont {P.}~\bibnamefont
  {Deuar}},\ }\href@noop {} {\enquote {\bibinfo {title} {The soliton regime in
  ultracold bose gases},}\ } (\bibinfo {year} {2017}),\ \bibinfo {note} {in
  preparation}\BibitemShut {NoStop}%
\bibitem [{\citenamefont {Sykes}\ \emph {et~al.}(2008)\citenamefont {Sykes},
  \citenamefont {Gangardt}, \citenamefont {Davis}, \citenamefont {Viering},
  \citenamefont {Raizen},\ and\ \citenamefont {Kheruntsyan}}]{Sykes08}%
  \BibitemOpen
  \bibfield  {author} {\bibinfo {author} {\bibfnamefont {A.~G.}\ \bibnamefont
  {Sykes}}, \bibinfo {author} {\bibfnamefont {D.~M.}\ \bibnamefont {Gangardt}},
  \bibinfo {author} {\bibfnamefont {M.~J.}\ \bibnamefont {Davis}}, \bibinfo
  {author} {\bibfnamefont {K.}~\bibnamefont {Viering}}, \bibinfo {author}
  {\bibfnamefont {M.~G.}\ \bibnamefont {Raizen}}, \ and\ \bibinfo {author}
  {\bibfnamefont {K.~V.}\ \bibnamefont {Kheruntsyan}},\ }\href {\doibase
  10.1103/PhysRevLett.100.160406} {\bibfield  {journal} {\bibinfo  {journal}
  {Phys. Rev. Lett.}\ }\textbf {\bibinfo {volume} {100}},\ \bibinfo {pages}
  {160406} (\bibinfo {year} {2008})}\BibitemShut {NoStop}%
\bibitem [{\citenamefont {Deuar}\ \emph {et~al.}(2009)\citenamefont {Deuar},
  \citenamefont {Sykes}, \citenamefont {Gangardt}, \citenamefont {Davis},
  \citenamefont {Drummond},\ and\ \citenamefont {Kheruntsyan}}]{Deuar09}%
  \BibitemOpen
  \bibfield  {author} {\bibinfo {author} {\bibfnamefont {P.}~\bibnamefont
  {Deuar}}, \bibinfo {author} {\bibfnamefont {A.~G.}\ \bibnamefont {Sykes}},
  \bibinfo {author} {\bibfnamefont {D.~M.}\ \bibnamefont {Gangardt}}, \bibinfo
  {author} {\bibfnamefont {M.~J.}\ \bibnamefont {Davis}}, \bibinfo {author}
  {\bibfnamefont {P.~D.}\ \bibnamefont {Drummond}}, \ and\ \bibinfo {author}
  {\bibfnamefont {K.~V.}\ \bibnamefont {Kheruntsyan}},\ }\href {\doibase
  10.1103/PhysRevA.79.043619} {\bibfield  {journal} {\bibinfo  {journal} {Phys.
  Rev. A}\ }\textbf {\bibinfo {volume} {79}},\ \bibinfo {pages} {043619}
  (\bibinfo {year} {2009})}\BibitemShut {NoStop}%
\bibitem [{\citenamefont {Wang}\ \emph {et~al.}(2013)\citenamefont {Wang},
  \citenamefont {Huang}, \citenamefont {Lee}, \citenamefont {Yin},
  \citenamefont {Guan},\ and\ \citenamefont {Batchelor}}]{Wang13}%
  \BibitemOpen
  \bibfield  {author} {\bibinfo {author} {\bibfnamefont {M.-S.}\ \bibnamefont
  {Wang}}, \bibinfo {author} {\bibfnamefont {J.-H.}\ \bibnamefont {Huang}},
  \bibinfo {author} {\bibfnamefont {C.-H.}\ \bibnamefont {Lee}}, \bibinfo
  {author} {\bibfnamefont {X.-G.}\ \bibnamefont {Yin}}, \bibinfo {author}
  {\bibfnamefont {X.-W.}\ \bibnamefont {Guan}}, \ and\ \bibinfo {author}
  {\bibfnamefont {M.~T.}\ \bibnamefont {Batchelor}},\ }\href {\doibase
  10.1103/PhysRevA.87.043634} {\bibfield  {journal} {\bibinfo  {journal} {Phys.
  Rev. A}\ }\textbf {\bibinfo {volume} {87}},\ \bibinfo {pages} {043634}
  (\bibinfo {year} {2013})}\BibitemShut {NoStop}%
\bibitem [{\citenamefont {Nandani}\ \emph {et~al.}(2016)\citenamefont
  {Nandani}, \citenamefont {Römer}, \citenamefont {Tan},\ and\ \citenamefont
  {Guan}}]{Nandani16}%
  \BibitemOpen
  \bibfield  {author} {\bibinfo {author} {\bibfnamefont {E.}~\bibnamefont
  {Nandani}}, \bibinfo {author} {\bibfnamefont {R.~A.}\ \bibnamefont {Römer}},
  \bibinfo {author} {\bibfnamefont {S.}~\bibnamefont {Tan}}, \ and\ \bibinfo
  {author} {\bibfnamefont {X.-W.}\ \bibnamefont {Guan}},\ }\href
  {http://stacks.iop.org/1367-2630/18/i=5/a=055014} {\bibfield  {journal}
  {\bibinfo  {journal} {New Journal of Physics}\ }\textbf {\bibinfo {volume}
  {18}},\ \bibinfo {pages} {055014} (\bibinfo {year} {2016})}\BibitemShut
  {NoStop}%
\bibitem [{\citenamefont {Lieb}\ and\ \citenamefont {Liniger}(1963)}]{Lieb63}%
  \BibitemOpen
  \bibfield  {author} {\bibinfo {author} {\bibfnamefont {E.~H.}\ \bibnamefont
  {Lieb}}\ and\ \bibinfo {author} {\bibfnamefont {W.}~\bibnamefont {Liniger}},\
  }\href {\doibase 10.1103/PhysRev.130.1605} {\bibfield  {journal} {\bibinfo
  {journal} {Phys. Rev.}\ }\textbf {\bibinfo {volume} {130}},\ \bibinfo {pages}
  {1605} (\bibinfo {year} {1963})}\BibitemShut {NoStop}%
\bibitem [{\citenamefont {Lieb}(1963)}]{Lieb63b}%
  \BibitemOpen
  \bibfield  {author} {\bibinfo {author} {\bibfnamefont {E.~H.}\ \bibnamefont
  {Lieb}},\ }\href {\doibase 10.1103/PhysRev.130.1616} {\bibfield  {journal}
  {\bibinfo  {journal} {Phys. Rev.}\ }\textbf {\bibinfo {volume} {130}},\
  \bibinfo {pages} {1616} (\bibinfo {year} {1963})}\BibitemShut {NoStop}%
\bibitem [{\citenamefont {Yang}\ and\ \citenamefont {Yang}(1969)}]{Yang69}%
  \BibitemOpen
  \bibfield  {author} {\bibinfo {author} {\bibfnamefont {C.~N.}\ \bibnamefont
  {Yang}}\ and\ \bibinfo {author} {\bibfnamefont {C.~P.}\ \bibnamefont
  {Yang}},\ }\href {\doibase http://dx.doi.org/10.1063/1.1664947} {\bibfield
  {journal} {\bibinfo  {journal} {Journal of Mathematical Physics}\ }\textbf
  {\bibinfo {volume} {10}},\ \bibinfo {pages} {1115} (\bibinfo {year}
  {1969})}\BibitemShut {NoStop}%
\bibitem [{\citenamefont {Kheruntsyan}\ \emph {et~al.}(2003)\citenamefont
  {Kheruntsyan}, \citenamefont {Gangardt}, \citenamefont {Drummond},\ and\
  \citenamefont {Shlyapnikov}}]{Kheruntsyan03}%
  \BibitemOpen
  \bibfield  {author} {\bibinfo {author} {\bibfnamefont {K.~V.}\ \bibnamefont
  {Kheruntsyan}}, \bibinfo {author} {\bibfnamefont {D.~M.}\ \bibnamefont
  {Gangardt}}, \bibinfo {author} {\bibfnamefont {P.~D.}\ \bibnamefont
  {Drummond}}, \ and\ \bibinfo {author} {\bibfnamefont {G.~V.}\ \bibnamefont
  {Shlyapnikov}},\ }\href {\doibase 10.1103/PhysRevLett.91.040403} {\bibfield
  {journal} {\bibinfo  {journal} {Phys. Rev. Lett.}\ }\textbf {\bibinfo
  {volume} {91}},\ \bibinfo {pages} {040403} (\bibinfo {year}
  {2003})}\BibitemShut {NoStop}%
\bibitem [{\citenamefont {Kheruntsyan}\ \emph {et~al.}(2005)\citenamefont
  {Kheruntsyan}, \citenamefont {Gangardt}, \citenamefont {Drummond},\ and\
  \citenamefont {Shlyapnikov}}]{Kheruntsyan05}%
  \BibitemOpen
  \bibfield  {author} {\bibinfo {author} {\bibfnamefont {K.~V.}\ \bibnamefont
  {Kheruntsyan}}, \bibinfo {author} {\bibfnamefont {D.~M.}\ \bibnamefont
  {Gangardt}}, \bibinfo {author} {\bibfnamefont {P.~D.}\ \bibnamefont
  {Drummond}}, \ and\ \bibinfo {author} {\bibfnamefont {G.~V.}\ \bibnamefont
  {Shlyapnikov}},\ }\href {\doibase 10.1103/PhysRevA.71.053615} {\bibfield
  {journal} {\bibinfo  {journal} {Phys. Rev. A}\ }\textbf {\bibinfo {volume}
  {71}},\ \bibinfo {pages} {053615} (\bibinfo {year} {2005})}\BibitemShut
  {NoStop}%
\bibitem [{\citenamefont {Cheianov}\ \emph {et~al.}(2006)\citenamefont
  {Cheianov}, \citenamefont {Smith},\ and\ \citenamefont
  {Zvonarev}}]{Cheianov06}%
  \BibitemOpen
  \bibfield  {author} {\bibinfo {author} {\bibfnamefont {V.~V.}\ \bibnamefont
  {Cheianov}}, \bibinfo {author} {\bibfnamefont {H.}~\bibnamefont {Smith}}, \
  and\ \bibinfo {author} {\bibfnamefont {M.~B.}\ \bibnamefont {Zvonarev}},\
  }\href {\doibase 10.1103/PhysRevA.73.051604} {\bibfield  {journal} {\bibinfo
  {journal} {Phys. Rev. A}\ }\textbf {\bibinfo {volume} {73}},\ \bibinfo
  {pages} {051604} (\bibinfo {year} {2006})}\BibitemShut {NoStop}%
\bibitem [{\citenamefont {Kormos}\ \emph {et~al.}(2011)\citenamefont {Kormos},
  \citenamefont {Chou},\ and\ \citenamefont {Imambekov}}]{Kormos11}%
  \BibitemOpen
  \bibfield  {author} {\bibinfo {author} {\bibfnamefont {M.}~\bibnamefont
  {Kormos}}, \bibinfo {author} {\bibfnamefont {Y.-Z.}\ \bibnamefont {Chou}}, \
  and\ \bibinfo {author} {\bibfnamefont {A.}~\bibnamefont {Imambekov}},\ }\href
  {\doibase 10.1103/PhysRevLett.107.230405} {\bibfield  {journal} {\bibinfo
  {journal} {Phys. Rev. Lett.}\ }\textbf {\bibinfo {volume} {107}},\ \bibinfo
  {pages} {230405} (\bibinfo {year} {2011})}\BibitemShut {NoStop}%
\bibitem [{\citenamefont {Kormos}\ \emph {et~al.}(2010)\citenamefont {Kormos},
  \citenamefont {Mussardo},\ and\ \citenamefont {Trombettoni}}]{Kormos10}%
  \BibitemOpen
  \bibfield  {author} {\bibinfo {author} {\bibfnamefont {M.}~\bibnamefont
  {Kormos}}, \bibinfo {author} {\bibfnamefont {G.}~\bibnamefont {Mussardo}}, \
  and\ \bibinfo {author} {\bibfnamefont {A.}~\bibnamefont {Trombettoni}},\
  }\href {\doibase 10.1103/PhysRevA.81.043606} {\bibfield  {journal} {\bibinfo
  {journal} {Phys. Rev. A}\ }\textbf {\bibinfo {volume} {81}},\ \bibinfo
  {pages} {043606} (\bibinfo {year} {2010})}\BibitemShut {NoStop}%
\bibitem [{\citenamefont {Piroli}\ and\ \citenamefont
  {Calabrese}(2015)}]{Piroli15}%
  \BibitemOpen
  \bibfield  {author} {\bibinfo {author} {\bibfnamefont {L.}~\bibnamefont
  {Piroli}}\ and\ \bibinfo {author} {\bibfnamefont {P.}~\bibnamefont
  {Calabrese}},\ }\href {http://stacks.iop.org/1751-8121/48/i=45/a=454002}
  {\bibfield  {journal} {\bibinfo  {journal} {Journal of Physics A:
  Mathematical and Theoretical}\ }\textbf {\bibinfo {volume} {48}},\ \bibinfo
  {pages} {454002} (\bibinfo {year} {2015})}\BibitemShut {NoStop}%
\bibitem [{\citenamefont {Panfil}\ and\ \citenamefont {Caux}(2014)}]{Panfil14}%
  \BibitemOpen
  \bibfield  {author} {\bibinfo {author} {\bibfnamefont {M.}~\bibnamefont
  {Panfil}}\ and\ \bibinfo {author} {\bibfnamefont {J.-S.}\ \bibnamefont
  {Caux}},\ }\href {\doibase 10.1103/PhysRevA.89.033605} {\bibfield  {journal}
  {\bibinfo  {journal} {Phys. Rev. A}\ }\textbf {\bibinfo {volume} {89}},\
  \bibinfo {pages} {033605} (\bibinfo {year} {2014})}\BibitemShut {NoStop}%
\bibitem [{\citenamefont {Panfil}\ \emph {et~al.}(2013)\citenamefont {Panfil},
  \citenamefont {De~Nardis},\ and\ \citenamefont {Caux}}]{Panfil13}%
  \BibitemOpen
  \bibfield  {author} {\bibinfo {author} {\bibfnamefont {M.}~\bibnamefont
  {Panfil}}, \bibinfo {author} {\bibfnamefont {J.}~\bibnamefont {De~Nardis}}, \
  and\ \bibinfo {author} {\bibfnamefont {J.-S.}\ \bibnamefont {Caux}},\ }\href
  {\doibase 10.1103/PhysRevLett.110.125302} {\bibfield  {journal} {\bibinfo
  {journal} {Phys. Rev. Lett.}\ }\textbf {\bibinfo {volume} {110}},\ \bibinfo
  {pages} {125302} (\bibinfo {year} {2013})}\BibitemShut {NoStop}%
\bibitem [{\citenamefont {De~Nardis}\ and\ \citenamefont
  {Panfil}(2016)}]{deNardis16}%
  \BibitemOpen
  \bibfield  {author} {\bibinfo {author} {\bibfnamefont {J.}~\bibnamefont
  {De~Nardis}}\ and\ \bibinfo {author} {\bibfnamefont {M.}~\bibnamefont
  {Panfil}},\ }\href {\doibase 10.21468/SciPostPhys.1.2.015} {\bibfield
  {journal} {\bibinfo  {journal} {SciPost Phys.}\ }\textbf {\bibinfo {volume}
  {1}},\ \bibinfo {pages} {015} (\bibinfo {year} {2016})}\BibitemShut {NoStop}%
\bibitem [{\citenamefont {Cherny}\ and\ \citenamefont
  {Brand}(2006)}]{Cherny06}%
  \BibitemOpen
  \bibfield  {author} {\bibinfo {author} {\bibfnamefont {A.~Y.}\ \bibnamefont
  {Cherny}}\ and\ \bibinfo {author} {\bibfnamefont {J.}~\bibnamefont {Brand}},\
  }\href {\doibase 10.1103/PhysRevA.73.023612} {\bibfield  {journal} {\bibinfo
  {journal} {Phys. Rev. A}\ }\textbf {\bibinfo {volume} {73}},\ \bibinfo
  {pages} {023612} (\bibinfo {year} {2006})}\BibitemShut {NoStop}%
\bibitem [{\citenamefont {Golovach}\ \emph {et~al.}(2009)\citenamefont
  {Golovach}, \citenamefont {Minguzzi},\ and\ \citenamefont
  {Glazman}}]{Golovach09}%
  \BibitemOpen
  \bibfield  {author} {\bibinfo {author} {\bibfnamefont {V.~N.}\ \bibnamefont
  {Golovach}}, \bibinfo {author} {\bibfnamefont {A.}~\bibnamefont {Minguzzi}},
  \ and\ \bibinfo {author} {\bibfnamefont {L.~I.}\ \bibnamefont {Glazman}},\
  }\href {\doibase 10.1103/PhysRevA.80.043611} {\bibfield  {journal} {\bibinfo
  {journal} {Phys. Rev. A}\ }\textbf {\bibinfo {volume} {80}},\ \bibinfo
  {pages} {043611} (\bibinfo {year} {2009})}\BibitemShut {NoStop}%
\bibitem [{\citenamefont {Kozlowski}\ \emph
  {et~al.}(2011{\natexlab{a}})\citenamefont {Kozlowski}, \citenamefont
  {Maillet},\ and\ \citenamefont {Slavnov}}]{Kozlowski11}%
  \BibitemOpen
  \bibfield  {author} {\bibinfo {author} {\bibfnamefont {K.~K.}\ \bibnamefont
  {Kozlowski}}, \bibinfo {author} {\bibfnamefont {J.~M.}\ \bibnamefont
  {Maillet}}, \ and\ \bibinfo {author} {\bibfnamefont {N.~A.}\ \bibnamefont
  {Slavnov}},\ }\href {http://stacks.iop.org/1742-5468/2011/i=03/a=P03019}
  {\bibfield  {journal} {\bibinfo  {journal} {Journal of Statistical Mechanics:
  Theory and Experiment}\ }\textbf {\bibinfo {volume} {2011}},\ \bibinfo
  {pages} {P03019} (\bibinfo {year} {2011}{\natexlab{a}})}\BibitemShut
  {NoStop}%
\bibitem [{\citenamefont {Kozlowski}\ \emph
  {et~al.}(2011{\natexlab{b}})\citenamefont {Kozlowski}, \citenamefont
  {Maillet},\ and\ \citenamefont {Slavnov}}]{Kozlowski11b}%
  \BibitemOpen
  \bibfield  {author} {\bibinfo {author} {\bibfnamefont {K.~K.}\ \bibnamefont
  {Kozlowski}}, \bibinfo {author} {\bibfnamefont {J.~M.}\ \bibnamefont
  {Maillet}}, \ and\ \bibinfo {author} {\bibfnamefont {N.~A.}\ \bibnamefont
  {Slavnov}},\ }\href {http://stacks.iop.org/1742-5468/2011/i=03/a=P03018}
  {\bibfield  {journal} {\bibinfo  {journal} {Journal of Statistical Mechanics:
  Theory and Experiment}\ }\textbf {\bibinfo {volume} {2011}},\ \bibinfo
  {pages} {P03018} (\bibinfo {year} {2011}{\natexlab{b}})}\BibitemShut
  {NoStop}%
\bibitem [{\citenamefont {P\^a\ifmmode~\mbox{\c{t}}\else \c{t}\fi{}u}\ and\
  \citenamefont {Kl\"umper}(2013)}]{Patu13}%
  \BibitemOpen
  \bibfield  {author} {\bibinfo {author} {\bibfnamefont {O.~I.}\ \bibnamefont
  {P\^a\ifmmode~\mbox{\c{t}}\else \c{t}\fi{}u}}\ and\ \bibinfo {author}
  {\bibfnamefont {A.}~\bibnamefont {Kl\"umper}},\ }\href {\doibase
  10.1103/PhysRevA.88.033623} {\bibfield  {journal} {\bibinfo  {journal} {Phys.
  Rev. A}\ }\textbf {\bibinfo {volume} {88}},\ \bibinfo {pages} {033623}
  (\bibinfo {year} {2013})}\BibitemShut {NoStop}%
\bibitem [{\citenamefont {Kl\"umper}\ and\ \citenamefont
  {P\^a\ifmmode~\mbox{\c{t}}\else \c{t}\fi{}u}(2014)}]{Klumper14}%
  \BibitemOpen
  \bibfield  {author} {\bibinfo {author} {\bibfnamefont {A.}~\bibnamefont
  {Kl\"umper}}\ and\ \bibinfo {author} {\bibfnamefont {O.~I.}\ \bibnamefont
  {P\^a\ifmmode~\mbox{\c{t}}\else \c{t}\fi{}u}},\ }\href {\doibase
  10.1103/PhysRevA.90.053626} {\bibfield  {journal} {\bibinfo  {journal} {Phys.
  Rev. A}\ }\textbf {\bibinfo {volume} {90}},\ \bibinfo {pages} {053626}
  (\bibinfo {year} {2014})}\BibitemShut {NoStop}%
\bibitem [{\citenamefont {Esteve}\ \emph {et~al.}(2006)\citenamefont {Esteve},
  \citenamefont {Trebbia}, \citenamefont {Schumm}, \citenamefont {Aspect},
  \citenamefont {Westbrook},\ and\ \citenamefont {Bouchoule}}]{Esteve06}%
  \BibitemOpen
  \bibfield  {author} {\bibinfo {author} {\bibfnamefont {J.}~\bibnamefont
  {Esteve}}, \bibinfo {author} {\bibfnamefont {J.-B.}\ \bibnamefont {Trebbia}},
  \bibinfo {author} {\bibfnamefont {T.}~\bibnamefont {Schumm}}, \bibinfo
  {author} {\bibfnamefont {A.}~\bibnamefont {Aspect}}, \bibinfo {author}
  {\bibfnamefont {C.~I.}\ \bibnamefont {Westbrook}}, \ and\ \bibinfo {author}
  {\bibfnamefont {I.}~\bibnamefont {Bouchoule}},\ }\href {\doibase
  10.1103/PhysRevLett.96.130403} {\bibfield  {journal} {\bibinfo  {journal}
  {Phys. Rev. Lett.}\ }\textbf {\bibinfo {volume} {96}},\ \bibinfo {pages}
  {130403} (\bibinfo {year} {2006})}\BibitemShut {NoStop}%
\bibitem [{\citenamefont {Sanner}\ \emph {et~al.}(2010)\citenamefont {Sanner},
  \citenamefont {Su}, \citenamefont {Keshet}, \citenamefont {Gommers},
  \citenamefont {Shin}, \citenamefont {Huang},\ and\ \citenamefont
  {Ketterle}}]{Sanner10}%
  \BibitemOpen
  \bibfield  {author} {\bibinfo {author} {\bibfnamefont {C.}~\bibnamefont
  {Sanner}}, \bibinfo {author} {\bibfnamefont {E.~J.}\ \bibnamefont {Su}},
  \bibinfo {author} {\bibfnamefont {A.}~\bibnamefont {Keshet}}, \bibinfo
  {author} {\bibfnamefont {R.}~\bibnamefont {Gommers}}, \bibinfo {author}
  {\bibfnamefont {Y.-i.}\ \bibnamefont {Shin}}, \bibinfo {author}
  {\bibfnamefont {W.}~\bibnamefont {Huang}}, \ and\ \bibinfo {author}
  {\bibfnamefont {W.}~\bibnamefont {Ketterle}},\ }\href {\doibase
  10.1103/PhysRevLett.105.040402} {\bibfield  {journal} {\bibinfo  {journal}
  {Phys. Rev. Lett.}\ }\textbf {\bibinfo {volume} {105}},\ \bibinfo {pages}
  {040402} (\bibinfo {year} {2010})}\BibitemShut {NoStop}%
\bibitem [{\citenamefont {M\"uller}\ \emph {et~al.}(2010)\citenamefont
  {M\"uller}, \citenamefont {Zimmermann}, \citenamefont {Meineke},
  \citenamefont {Brantut}, \citenamefont {Esslinger},\ and\ \citenamefont
  {Moritz}}]{Muller10}%
  \BibitemOpen
  \bibfield  {author} {\bibinfo {author} {\bibfnamefont {T.}~\bibnamefont
  {M\"uller}}, \bibinfo {author} {\bibfnamefont {B.}~\bibnamefont
  {Zimmermann}}, \bibinfo {author} {\bibfnamefont {J.}~\bibnamefont {Meineke}},
  \bibinfo {author} {\bibfnamefont {J.-P.}\ \bibnamefont {Brantut}}, \bibinfo
  {author} {\bibfnamefont {T.}~\bibnamefont {Esslinger}}, \ and\ \bibinfo
  {author} {\bibfnamefont {H.}~\bibnamefont {Moritz}},\ }\href {\doibase
  10.1103/PhysRevLett.105.040401} {\bibfield  {journal} {\bibinfo  {journal}
  {Phys. Rev. Lett.}\ }\textbf {\bibinfo {volume} {105}},\ \bibinfo {pages}
  {040401} (\bibinfo {year} {2010})}\BibitemShut {NoStop}%
\bibitem [{\citenamefont {Armijo}\ \emph {et~al.}(2010)\citenamefont {Armijo},
  \citenamefont {Jacqmin}, \citenamefont {Kheruntsyan},\ and\ \citenamefont
  {Bouchoule}}]{Armijo10}%
  \BibitemOpen
  \bibfield  {author} {\bibinfo {author} {\bibfnamefont {J.}~\bibnamefont
  {Armijo}}, \bibinfo {author} {\bibfnamefont {T.}~\bibnamefont {Jacqmin}},
  \bibinfo {author} {\bibfnamefont {K.~V.}\ \bibnamefont {Kheruntsyan}}, \ and\
  \bibinfo {author} {\bibfnamefont {I.}~\bibnamefont {Bouchoule}},\ }\href
  {\doibase 10.1103/PhysRevLett.105.230402} {\bibfield  {journal} {\bibinfo
  {journal} {Phys. Rev. Lett.}\ }\textbf {\bibinfo {volume} {105}},\ \bibinfo
  {pages} {230402} (\bibinfo {year} {2010})}\BibitemShut {NoStop}%
\bibitem [{\citenamefont {Jacqmin}\ \emph {et~al.}(2011)\citenamefont
  {Jacqmin}, \citenamefont {Armijo}, \citenamefont {Berrada}, \citenamefont
  {Kheruntsyan},\ and\ \citenamefont {Bouchoule}}]{Jacqmin11}%
  \BibitemOpen
  \bibfield  {author} {\bibinfo {author} {\bibfnamefont {T.}~\bibnamefont
  {Jacqmin}}, \bibinfo {author} {\bibfnamefont {J.}~\bibnamefont {Armijo}},
  \bibinfo {author} {\bibfnamefont {T.}~\bibnamefont {Berrada}}, \bibinfo
  {author} {\bibfnamefont {K.~V.}\ \bibnamefont {Kheruntsyan}}, \ and\ \bibinfo
  {author} {\bibfnamefont {I.}~\bibnamefont {Bouchoule}},\ }\href {\doibase
  10.1103/PhysRevLett.106.230405} {\bibfield  {journal} {\bibinfo  {journal}
  {Phys. Rev. Lett.}\ }\textbf {\bibinfo {volume} {106}},\ \bibinfo {pages}
  {230405} (\bibinfo {year} {2011})}\BibitemShut {NoStop}%
\bibitem [{\citenamefont {Armijo}\ \emph {et~al.}(2011)\citenamefont {Armijo},
  \citenamefont {Jacqmin}, \citenamefont {Kheruntsyan},\ and\ \citenamefont
  {Bouchoule}}]{Armijo11}%
  \BibitemOpen
  \bibfield  {author} {\bibinfo {author} {\bibfnamefont {J.}~\bibnamefont
  {Armijo}}, \bibinfo {author} {\bibfnamefont {T.}~\bibnamefont {Jacqmin}},
  \bibinfo {author} {\bibfnamefont {K.}~\bibnamefont {Kheruntsyan}}, \ and\
  \bibinfo {author} {\bibfnamefont {I.}~\bibnamefont {Bouchoule}},\ }\href
  {\doibase 10.1103/PhysRevA.83.021605} {\bibfield  {journal} {\bibinfo
  {journal} {Phys. Rev. A}\ }\textbf {\bibinfo {volume} {83}},\ \bibinfo
  {pages} {021605} (\bibinfo {year} {2011})}\BibitemShut {NoStop}%
\bibitem [{\citenamefont {Armijo}(2012)}]{Armijo12}%
  \BibitemOpen
  \bibfield  {author} {\bibinfo {author} {\bibfnamefont {J.}~\bibnamefont
  {Armijo}},\ }\href {\doibase 10.1103/PhysRevLett.108.225306} {\bibfield
  {journal} {\bibinfo  {journal} {Phys. Rev. Lett.}\ }\textbf {\bibinfo
  {volume} {108}},\ \bibinfo {pages} {225306} (\bibinfo {year}
  {2012})}\BibitemShut {NoStop}%
\bibitem [{\citenamefont {Bisset}\ \emph {et~al.}(2013)\citenamefont {Bisset},
  \citenamefont {Ticknor},\ and\ \citenamefont {Blakie}}]{Bisset13}%
  \BibitemOpen
  \bibfield  {author} {\bibinfo {author} {\bibfnamefont {R.~N.}\ \bibnamefont
  {Bisset}}, \bibinfo {author} {\bibfnamefont {C.}~\bibnamefont {Ticknor}}, \
  and\ \bibinfo {author} {\bibfnamefont {P.~B.}\ \bibnamefont {Blakie}},\
  }\href {\doibase 10.1103/PhysRevA.88.063624} {\bibfield  {journal} {\bibinfo
  {journal} {Phys. Rev. A}\ }\textbf {\bibinfo {volume} {88}},\ \bibinfo
  {pages} {063624} (\bibinfo {year} {2013})}\BibitemShut {NoStop}%
\bibitem [{\citenamefont {Henkel}\ \emph {et~al.}(2017)\citenamefont {Henkel},
  \citenamefont {Sauer},\ and\ \citenamefont {Proukakis}}]{Henkel17}%
  \BibitemOpen
  \bibfield  {author} {\bibinfo {author} {\bibfnamefont {C.}~\bibnamefont
  {Henkel}}, \bibinfo {author} {\bibfnamefont {T.-O.}\ \bibnamefont {Sauer}}, \
  and\ \bibinfo {author} {\bibfnamefont {N.~P.}\ \bibnamefont {Proukakis}},\
  }\href@noop {} {\bibfield  {journal} {\bibinfo  {journal} {arXiv:1701.03133}\
  } (\bibinfo {year} {2017})}\BibitemShut {NoStop}%
\bibitem [{\citenamefont {Perrin}\ \emph {et~al.}(2007)\citenamefont {Perrin},
  \citenamefont {Chang}, \citenamefont {Krachmalnicoff}, \citenamefont
  {Schellekens}, \citenamefont {Boiron}, \citenamefont {Aspect},\ and\
  \citenamefont {Westbrook}}]{Perrin07}%
  \BibitemOpen
  \bibfield  {author} {\bibinfo {author} {\bibfnamefont {A.}~\bibnamefont
  {Perrin}}, \bibinfo {author} {\bibfnamefont {H.}~\bibnamefont {Chang}},
  \bibinfo {author} {\bibfnamefont {V.}~\bibnamefont {Krachmalnicoff}},
  \bibinfo {author} {\bibfnamefont {M.}~\bibnamefont {Schellekens}}, \bibinfo
  {author} {\bibfnamefont {D.}~\bibnamefont {Boiron}}, \bibinfo {author}
  {\bibfnamefont {A.}~\bibnamefont {Aspect}}, \ and\ \bibinfo {author}
  {\bibfnamefont {C.~I.}\ \bibnamefont {Westbrook}},\ }\href {\doibase
  10.1103/PhysRevLett.99.150405} {\bibfield  {journal} {\bibinfo  {journal}
  {Phys. Rev. Lett.}\ }\textbf {\bibinfo {volume} {99}},\ \bibinfo {pages}
  {150405} (\bibinfo {year} {2007})}\BibitemShut {NoStop}%
\bibitem [{\citenamefont {Jaskula}\ \emph {et~al.}(2010)\citenamefont
  {Jaskula}, \citenamefont {Bonneau}, \citenamefont {Partridge}, \citenamefont
  {Krachmalnicoff}, \citenamefont {Deuar}, \citenamefont {Kheruntsyan},
  \citenamefont {Aspect}, \citenamefont {Boiron},\ and\ \citenamefont
  {Westbrook}}]{Jaskula10}%
  \BibitemOpen
  \bibfield  {author} {\bibinfo {author} {\bibfnamefont {J.-C.}\ \bibnamefont
  {Jaskula}}, \bibinfo {author} {\bibfnamefont {M.}~\bibnamefont {Bonneau}},
  \bibinfo {author} {\bibfnamefont {G.~B.}\ \bibnamefont {Partridge}}, \bibinfo
  {author} {\bibfnamefont {V.}~\bibnamefont {Krachmalnicoff}}, \bibinfo
  {author} {\bibfnamefont {P.}~\bibnamefont {Deuar}}, \bibinfo {author}
  {\bibfnamefont {K.~V.}\ \bibnamefont {Kheruntsyan}}, \bibinfo {author}
  {\bibfnamefont {A.}~\bibnamefont {Aspect}}, \bibinfo {author} {\bibfnamefont
  {D.}~\bibnamefont {Boiron}}, \ and\ \bibinfo {author} {\bibfnamefont {C.~I.}\
  \bibnamefont {Westbrook}},\ }\href {\doibase 10.1103/PhysRevLett.105.190402}
  {\bibfield  {journal} {\bibinfo  {journal} {Phys. Rev. Lett.}\ }\textbf
  {\bibinfo {volume} {105}},\ \bibinfo {pages} {190402} (\bibinfo {year}
  {2010})}\BibitemShut {NoStop}%
\bibitem [{\citenamefont {Kheruntsyan}\ \emph {et~al.}(2012)\citenamefont
  {Kheruntsyan}, \citenamefont {Jaskula}, \citenamefont {Deuar}, \citenamefont
  {Bonneau}, \citenamefont {Partridge}, \citenamefont {Ruaudel}, \citenamefont
  {Lopes}, \citenamefont {Boiron},\ and\ \citenamefont
  {Westbrook}}]{Kheruntsyan12}%
  \BibitemOpen
  \bibfield  {author} {\bibinfo {author} {\bibfnamefont {K.~V.}\ \bibnamefont
  {Kheruntsyan}}, \bibinfo {author} {\bibfnamefont {J.-C.}\ \bibnamefont
  {Jaskula}}, \bibinfo {author} {\bibfnamefont {P.}~\bibnamefont {Deuar}},
  \bibinfo {author} {\bibfnamefont {M.}~\bibnamefont {Bonneau}}, \bibinfo
  {author} {\bibfnamefont {G.~B.}\ \bibnamefont {Partridge}}, \bibinfo {author}
  {\bibfnamefont {J.}~\bibnamefont {Ruaudel}}, \bibinfo {author} {\bibfnamefont
  {R.}~\bibnamefont {Lopes}}, \bibinfo {author} {\bibfnamefont
  {D.}~\bibnamefont {Boiron}}, \ and\ \bibinfo {author} {\bibfnamefont {C.~I.}\
  \bibnamefont {Westbrook}},\ }\href {\doibase 10.1103/PhysRevLett.108.260401}
  {\bibfield  {journal} {\bibinfo  {journal} {Phys. Rev. Lett.}\ }\textbf
  {\bibinfo {volume} {108}},\ \bibinfo {pages} {260401} (\bibinfo {year}
  {2012})}\BibitemShut {NoStop}%
\bibitem [{\citenamefont {Dall}\ \emph {et~al.}(2013)\citenamefont {Dall},
  \citenamefont {Manning}, \citenamefont {Hodgman}, \citenamefont {RuGway},
  \citenamefont {Kheruntsyan},\ and\ \citenamefont {Truscott}}]{Dall13}%
  \BibitemOpen
  \bibfield  {author} {\bibinfo {author} {\bibfnamefont {R.~G.}\ \bibnamefont
  {Dall}}, \bibinfo {author} {\bibfnamefont {A.~G.}\ \bibnamefont {Manning}},
  \bibinfo {author} {\bibfnamefont {S.~S.}\ \bibnamefont {Hodgman}}, \bibinfo
  {author} {\bibfnamefont {W.}~\bibnamefont {RuGway}}, \bibinfo {author}
  {\bibfnamefont {K.~V.}\ \bibnamefont {Kheruntsyan}}, \ and\ \bibinfo {author}
  {\bibfnamefont {A.~G.}\ \bibnamefont {Truscott}},\ }\href {\doibase
  10.1038/nphys2632} {\bibfield  {journal} {\bibinfo  {journal} {Nature
  Physics}\ }\textbf {\bibinfo {volume} {9}},\ \bibinfo {pages} {341} (\bibinfo
  {year} {2013})}\BibitemShut {NoStop}%
\bibitem [{\citenamefont {Manning}\ \emph {et~al.}(2013)\citenamefont
  {Manning}, \citenamefont {RuGway}, \citenamefont {Hodgman}, \citenamefont
  {Dall}, \citenamefont {Baldwin},\ and\ \citenamefont {Truscott}}]{Manning13}%
  \BibitemOpen
  \bibfield  {author} {\bibinfo {author} {\bibfnamefont {A.~G.}\ \bibnamefont
  {Manning}}, \bibinfo {author} {\bibfnamefont {W.}~\bibnamefont {RuGway}},
  \bibinfo {author} {\bibfnamefont {S.~S.}\ \bibnamefont {Hodgman}}, \bibinfo
  {author} {\bibfnamefont {R.~G.}\ \bibnamefont {Dall}}, \bibinfo {author}
  {\bibfnamefont {K.~G.~H.}\ \bibnamefont {Baldwin}}, \ and\ \bibinfo {author}
  {\bibfnamefont {A.~G.}\ \bibnamefont {Truscott}},\ }\href
  {http://stacks.iop.org/1367-2630/15/i=1/a=013042} {\bibfield  {journal}
  {\bibinfo  {journal} {New Journal of Physics}\ }\textbf {\bibinfo {volume}
  {15}},\ \bibinfo {pages} {013042} (\bibinfo {year} {2013})}\BibitemShut
  {NoStop}%
\bibitem [{\citenamefont {Carusotto}\ and\ \citenamefont
  {Castin}(2001)}]{Carusotto01}%
  \BibitemOpen
  \bibfield  {author} {\bibinfo {author} {\bibfnamefont {I.}~\bibnamefont
  {Carusotto}}\ and\ \bibinfo {author} {\bibfnamefont {Y.}~\bibnamefont
  {Castin}},\ }\href {http://stacks.iop.org/0953-4075/34/i=23/a=305} {\bibfield
   {journal} {\bibinfo  {journal} {Journal of Physics B: Atomic, Molecular and
  Optical Physics}\ }\textbf {\bibinfo {volume} {34}},\ \bibinfo {pages} {4589}
  (\bibinfo {year} {2001})}\BibitemShut {NoStop}%
\bibitem [{\citenamefont {Mora}\ and\ \citenamefont {Castin}(2003)}]{Mora03}%
  \BibitemOpen
  \bibfield  {author} {\bibinfo {author} {\bibfnamefont {C.}~\bibnamefont
  {Mora}}\ and\ \bibinfo {author} {\bibfnamefont {Y.}~\bibnamefont {Castin}},\
  }\href {\doibase 10.1103/PhysRevA.67.053615} {\bibfield  {journal} {\bibinfo
  {journal} {Phys. Rev. A}\ }\textbf {\bibinfo {volume} {67}},\ \bibinfo
  {pages} {053615} (\bibinfo {year} {2003})}\BibitemShut {NoStop}%
\bibitem [{\citenamefont {Pietraszewicz}\ and\ \citenamefont
  {Deuar}(2017)}]{Pietraszewicz17}%
  \BibitemOpen
  \bibfield  {author} {\bibinfo {author} {\bibfnamefont {J.}~\bibnamefont
  {Pietraszewicz}}\ and\ \bibinfo {author} {\bibfnamefont {P.}~\bibnamefont
  {Deuar}},\ }\href@noop {} {\bibfield  {journal} {\bibinfo  {journal}
  {arXiv:1707.01776}\ } (\bibinfo {year} {2017})}\BibitemShut {NoStop}%
\bibitem [{\citenamefont {Deuar}\ \emph {et~al.}(2013)\citenamefont {Deuar},
  \citenamefont {Wasak}, \citenamefont {Zi\'{n}}, \citenamefont
  {Chwede\'{n}czuk},\ and\ \citenamefont {Trippenbach}}]{Deuar13}%
  \BibitemOpen
  \bibfield  {author} {\bibinfo {author} {\bibfnamefont {P.}~\bibnamefont
  {Deuar}}, \bibinfo {author} {\bibfnamefont {T.}~\bibnamefont {Wasak}},
  \bibinfo {author} {\bibfnamefont {P.}~\bibnamefont {Zi\'{n}}}, \bibinfo
  {author} {\bibfnamefont {J.}~\bibnamefont {Chwede\'{n}czuk}}, \ and\ \bibinfo
  {author} {\bibfnamefont {M.}~\bibnamefont {Trippenbach}},\ }\href {\doibase
  10.1103/PhysRevA.88.013617} {\bibfield  {journal} {\bibinfo  {journal} {Phys.
  Rev. A}\ }\textbf {\bibinfo {volume} {88}},\ \bibinfo {pages} {013617}
  (\bibinfo {year} {2013})}\BibitemShut {NoStop}%
\bibitem [{\citenamefont {De~Rosi}\ \emph {et~al.}(2017)\citenamefont
  {De~Rosi}, \citenamefont {Astrakharchik},\ and\ \citenamefont
  {Stringari}}]{DeRosi17}%
  \BibitemOpen
  \bibfield  {author} {\bibinfo {author} {\bibfnamefont {G.}~\bibnamefont
  {De~Rosi}}, \bibinfo {author} {\bibfnamefont {G.~E.}\ \bibnamefont
  {Astrakharchik}}, \ and\ \bibinfo {author} {\bibfnamefont {S.}~\bibnamefont
  {Stringari}},\ }\href {\doibase 10.1103/PhysRevA.96.013613} {\bibfield
  {journal} {\bibinfo  {journal} {Phys. Rev. A}\ }\textbf {\bibinfo {volume}
  {96}},\ \bibinfo {pages} {013613} (\bibinfo {year} {2017})}\BibitemShut
  {NoStop}%
\bibitem [{\citenamefont {van Amerongen}\ \emph {et~al.}(2008)\citenamefont
  {van Amerongen}, \citenamefont {van Es}, \citenamefont {Wicke}, \citenamefont
  {Kheruntsyan},\ and\ \citenamefont {van Druten}}]{vanAmerongen08}%
  \BibitemOpen
  \bibfield  {author} {\bibinfo {author} {\bibfnamefont {A.~H.}\ \bibnamefont
  {van Amerongen}}, \bibinfo {author} {\bibfnamefont {J.~J.~P.}\ \bibnamefont
  {van Es}}, \bibinfo {author} {\bibfnamefont {P.}~\bibnamefont {Wicke}},
  \bibinfo {author} {\bibfnamefont {K.~V.}\ \bibnamefont {Kheruntsyan}}, \ and\
  \bibinfo {author} {\bibfnamefont {N.~J.}\ \bibnamefont {van Druten}},\ }\href
  {\doibase 10.1103/PhysRevLett.100.090402} {\bibfield  {journal} {\bibinfo
  {journal} {Phys. Rev. Lett.}\ }\textbf {\bibinfo {volume} {100}},\ \bibinfo
  {pages} {090402} (\bibinfo {year} {2008})}\BibitemShut {NoStop}%
\bibitem [{\citenamefont {Jacqmin}\ \emph {et~al.}(2012)\citenamefont
  {Jacqmin}, \citenamefont {Fang}, \citenamefont {Berrada}, \citenamefont
  {Roscilde},\ and\ \citenamefont {Bouchoule}}]{Jacqmin12}%
  \BibitemOpen
  \bibfield  {author} {\bibinfo {author} {\bibfnamefont {T.}~\bibnamefont
  {Jacqmin}}, \bibinfo {author} {\bibfnamefont {B.}~\bibnamefont {Fang}},
  \bibinfo {author} {\bibfnamefont {T.}~\bibnamefont {Berrada}}, \bibinfo
  {author} {\bibfnamefont {T.}~\bibnamefont {Roscilde}}, \ and\ \bibinfo
  {author} {\bibfnamefont {I.}~\bibnamefont {Bouchoule}},\ }\href {\doibase
  10.1103/PhysRevA.86.043626} {\bibfield  {journal} {\bibinfo  {journal} {Phys.
  Rev. A}\ }\textbf {\bibinfo {volume} {86}},\ \bibinfo {pages} {043626}
  (\bibinfo {year} {2012})}\BibitemShut {NoStop}%
\end{thebibliography}%

\end{document}